\documentclass[10pt]{article}

\usepackage{natbib}
\usepackage{enumerate,array}
\usepackage{mathrsfs} 
\usepackage{amsfonts, amssymb, amsmath,amsthm, slashbox}
\usepackage{cancel, url, graphicx, bm}
\usepackage{hyperref}
\usepackage{changes}  
\usepackage{comment}
\input{colordvi}
\def\em{\it}
\def\emph{\textit}

\newcommand{\newtag}{\tag{\theequation}\addtocounter{equation}{1}}
\def\o*{o_{{\!}_{P^*}}}
\def\O*{\cal O_{{\!}_{P^*}}}

\def\E{\mathrm{E}}
\def\Var{\mathrm{Var}}

\def\B{\mathrm{B}}

\def\.{\mbox{.}}

\def\diag{\mbox{diag}}

\def\sfrac(#1,#2){\mbox{$\frac{#1}{#2}$}}
\def\ints(#1,#2){\mathbb{I}_{#1}^{#2}}
\def\ie{\textit{i.e.}}
\def\ie{\textit{i.e. }}

\def\T{{\!\mathrm{\scriptscriptstyle \top}\!}}

\newtheorem {theorem}{Theorem}
\newtheorem {lemma}{Lemma}
\newtheorem {rem}{Remark}
\newtheorem{assump}{}

\newtheorem{cond}{}

\newtheorem{prop}[theorem]{Proposition}
\title{Maximum Approximate Bernstein Likelihood Estimation in Proportional Hazard Model for Interval-Censored Data} 
\author{Zhong Guan\\
Department of Mathematical Sciences\\
Indiana University South Bend, USA\\
zguan@iusb.edu\\
}
\date{}

\begin{document}
\maketitle
\section*{Abstract}
Maximum approximate Bernstein likelihood
estimates of the baseline density function and the regression coefficients in the
proportional hazard regression models based on interval-censored
event time data are proposed. This results in not only a smooth estimate of the
survival function which enjoys faster convergence rate but also improved estimates of
the regression coefficients. Simulation shows that the finite sample performance of the proposed
method is better than the existing ones. The proposed method is illustrated by real data
applications.

\noindent\textbf{Key Words and Phrases:} Approximate Likelihood, Bernstein Polynomial Model, Cox's Proportional Hazard Regression Model, Density Estimation,
Interval Censoring,  Survival Curve.
\section{Introduction}
\label{sect: introduction}
Traditionally in semi- and nonparametric statistics we approximate an unknown smooth distribution function by a step function and parameterize this infinite-dimensional parameter by the jump sizes of the step function at the observed values. Therefore, the working model is actually of finite but varying dimension.
The resulting estimate is a step function and does not deserve a density. This approach works fine when the infinite-dimensional parameter is nuisance. However, in the situation when such parameters such as survival, hazard, and density functions are our concerns the traditional approach which results in a jagged step-function estimation is not satisfactory especially when sample size is small which is usually the case for survival analysis of rare diseases. Besides the roughness of the estimation when data are incompletely observed it is difficult to parameterize the unknown survival function and not easy to find the nonparametric maximum likelihood estimate due to the complication of assigning probabilities and the large number of parameters (usually the same as the sample size) to be estimated. Moreover, the roughness of the estimate of nonparametric component could reduce the accuracy of the estimates of parameters in semiparametric models. \cite{Turnbull-1976-JRSSB} presented an EM algorithm \citep{Dempster1977} to compute the {\em discrete} nonparametric maximum likelihood estimate (NPMLE) of the distribution function from grouped, censored, and truncated data without covariates \citep[see also][]{Groeneboom-and-Wellner-1992-book}. The method is generalized to obtain semiparametric maximum likelihood estimate (SPMLE) of the survival function to models including Cox's proportional hazards (PH) model by \cite{Finkelstein-1986-biom}, \cite{Huang-1996-aos}, \cite{Huang-and-Wellner-1997}, and \cite{Pan-1999-JCGS}. \cite{Finkelstein-and-Wolfe-1985-biom} proposed some semiparametric models for interval censored data. Asymptotic results about some semiparametric models can be found in  \cite{Huang-and-Wellner-1997}, and  \cite{Schick-and-Yu-2000-sjs}, etc.  With interval censored data the assignment of the probabilities within the Turnbull interval cannot be uniquely determined \citep{Anderson-Bergman-2017-JSS}. \cite{Groeneboom-and-Wellner-1992-book} suggested an iterative convex minorant (ICM) algorithm, which was improved or generalized by
\cite{Wellner-and-Zhan-1997-JASA},  \cite{Pan-1999-JCGS}, and \cite{Anderson-Bergman-2017-JCGS}.  Grouped failure time data have been studied by, among others, \cite{Prentice-and-Gloeckler-1978-biom} and \cite{Pierce-Stewart-and-Kopecky-1979-biom}. Unfortunately, the NPMLE or SPMLE of the survival function
is a step-function and may be not unique.  Parametric models and Kernel {\em smoothing} methods \citep{Parzen-1962, Rosenblatt-1956-ann-math-stat} have been applied to obtain smooth estimator of survival function \citep{Lindsey-1998-LDA, Lindsey-1998-Stat-Med,Betensky-etal-1999-biom}.
Another continuous estimation was due to \cite{Becker-and-Melbye-1990-australian-J-Stat} who assumed piecewise
constant intensity model.
\cite{Carstensen-1996-SIM} generalized this method to regression models by assuming piecewise constant baseline rate.

\cite{Goetghebeur-and-Ryan-2000-biom} indicated that many of the EM-like methods have the relatively {\em ad hoc} nature of the procedure used to impute missing data and proposed a method using approximate likelihood to avoid such problem that retains some of the appealing features of the nonparametric smoothing methods such as the regression spline smoothing of \cite{Kooperberg-1997-biom} and the local likelihood kernel smoothing of \cite{Betensky-etal-1999-biom}.

Nonparametric density estimation is rather difficult due the lack of information contained in sample about it \citep{Bickel-et-al-book-1998,Ibragimov-Hasminskii-1982}. Kernel method is usually unsatisfactory when sample size is small
even for complete data. 
Some authors have studied the estimation of density function based on censored data \citep[see for example][and the refereces therein]{ Braun-etal-2005-cjs, Harlass-2016-phd-thesis} without covariate.

A useful {\em working statistical model} must be finite-dimensional and {\em approximates} \citep[see page 1 of][]{Bickel-et-al-book-1998} the true underlying distribution. Instead of approximating the underlying continuous distribution function by a step-function which is a multinomial probability model, \cite{Guan-jns-2015} 
suggested a Bernstein polynomial approximation \citep{Bernstein,Lorentz-1963-Math-Annalen} which is actually a mixture of some specific beta distributions. This Bernstein polynomial model performs much better than the classical kernel method  for estimating density even from grouped data \citep{Guan-2017-jns}. The maximum approximate Bernstein likelihood estimate  can be viewed as a {\em continuous} version of the NPMLE or SPMLE.
In this paper such  estimates of the conditional survival and density  functions given covariate are proposed by fitting interval censored data with Cox's proportional hazards  model.
\section{Methodology}

\subsection{Proportional Hazards Model}
Let $T$ be an event time and $\bm X$ be an associated $d$-dimensional covariate with distribution $H(\bm x)$ on $\cal{X}$. We denote the marginal and the conditional survival functions of $T$, respectively, by
 $S(t) = \bar F(t)= 1- F(t)=P(T > t)$ and $ S(t|\bm{x}) = \bar F(t|\bm{x})=  1- F(t|\bm{x})=P(T > t|\bm X=\bm{x}).$
Let $f(t|\bm{x})$ denote the conditional density of a continuous $T$ given $\bm X=\bm x$. The conditional cumulative hazard function, odds ratio, and hazard rate are, respectively,
 $$\Lambda(t|\bm x)=-\log S(t|\bm x), \quad O(y|\bm x)=\frac{S(y|\bm x)}{1-S(y|\bm x)},\quad \lambda(t|\bm x)=\frac{d}{dt}\Lambda(t|\bm x)=\frac{f(t|\bm x)}{S(t|\bm x)}.$$
 Consider the Cox's proportional hazard (PH) regression model \citep{Cox1972}
\begin{equation}\label{eq: Cox PH model-conditional survival function}
S(t|\bm x) =S(t|\bm x, \bm\gamma, f_0)=S(t|\bm x_0)^{\exp(\bm\gamma^\T\tilde{\bm x})},
\end{equation}
where $\bm\gamma\in \Gamma\subset \mathbb{R}^d$, $\tilde{\bm x} =\bm x -\bm x_0$, $\bm x_0$ is any fixed covariate value, $f_0(\cdot)=f(\cdot|\bm x_0)$ is the unknown baseline density and  $S(\cdot|\bm x_0)=\int_\cdot^\infty f(t|\bm x_0)dt$ is the corresponding survival function.
This 
is equivalent to
\begin{equation}\label{eq: Cox PH model-conditional density}
f(t|\bm x) = f(t|\bm x;\bm\gamma, f_0)={\exp(\bm\gamma^\T\tilde{\bm x})}S(t|\bm x_0)^{\exp(\bm\gamma^\T\tilde{\bm x})-1}f(t|\bm x_0).
\end{equation}
It is clear that  (\ref{eq: Cox PH model-conditional survival function}) and (\ref{eq: Cox PH model-conditional density}) are also true if we change the ``baseline'' covariate $\bm x_0$ to any $\bm x_0^*\in \cal{X}$ with the same $\bm\gamma$ but $\tilde{\bm x}$ being replaced by $\tilde{\bm x}^*=\bm x-\bm x_0^*$.
For a given $\bm \gamma\in\Gamma$, define a $\bm\gamma$-related ``baseline'' as an $\bm x_{\bm \gamma}\in\arg\min_{\bm x\in\mathcal{X}} \bm \gamma^\T\bm x $  and denote $\tilde{\bm x}_{\bm\gamma}=\bm x-\bm x_{\bm\gamma}$.
Define $\tau=\inf\{t: F(t|\bm x_{0})=1\}$.
It is true that $\tau$ is independent of $\bm x_0$, $0<\tau\le\infty$, and $f(t|\bm x)$ have the same support $[0,\tau]$ for all $\bm x\in \cal{X}$.
It is obvious that for any strictly increasing continuous function $\psi$, $P(\psi(T)> t|\bm x)=P(\psi(T)> t|\bm x_0)^{\exp(\bm\gamma^\T\tilde{\bm x})}$.
Thus the transformed event time $\psi(T)$ also satisfies the Cox model (\ref{eq: Cox PH model-conditional survival function}).


 We will consider the general situation where the event time is subject to interval censoring.  The observed data are $\bm Z=(\bm Y, \bm X,  \Delta)$, where $\bm Y=(Y_1,Y_2]$ and $\Delta$ is the censoring indicator, i.e., $T=Y=Y_1=Y_2$ is uncensored if $\Delta=0$ and $T\in\bm Y=(Y_1,Y_2]$, $0\le Y_1<Y_2\le\infty$, is interval censored if $\Delta=1$.
The reader is referred to \cite{Huang-and-Wellner-1997} for a review and more references about interval censoring.
%
The right-censoring $Y_2=\infty$
and left-censoring $Y_1=0$ are included as special cases.
For any individual observation $\bm z=(\bm y, \bm x,  \delta)$, where if $\delta=0$ then $\bm y=y=t$  else if $\delta=1$ then $\bm y=(y_1,y_2]\ni t$,  $0\le y_1<y_2\le \infty$, 
the full loglikelihood, up to an additive term independent of $(\bm\gamma, f_0)$, is
\begin{align}\nonumber
\ell(\bm\gamma, f_0;\bm z)
=&~ (1-\delta) [{\bm\gamma^\T \tilde{\bm x}} +\log f(y|\bm x_0)-(e^{\bm\gamma^\T \tilde{\bm x}}-1)\Lambda(y|\bm x_0)]
\\\label{eq: loglikelihood ell(gamma,p) for PH model}
& +  {\delta} \log[S(y_{1}|\bm x_0)^{e^{\bm\gamma^\T \tilde{\bm x}}}-S(y_{2}|\bm x_0)^{e^{\bm\gamma^\T \tilde{\bm x}}}].
\end{align}

Let $(\bm y_i, \bm x_i, \delta_i)$, $i\in\ints(1,n)$ be independent observations of $(\bm Y, \bm X,  \Delta)$, here  and in what follows $\mathbb{I}_m^n=\{m,\ldots,n\}$ for any integers $m\le n\le\infty$. 
If $\tau$ is either unknown or $\tau=\infty$ and $\tau_n$  is at least the last finite observed time, i.e., $\tau_n\ge y_{(n)}=\max\{y_{i1}, y_{j2}: y_{j2}<\infty;\, i,j\in\ints(1,n)\}$ then $[\tau_n, \infty)$  is contained in the last {\em Turnbull interval} \citep{Turnbull-1976-JRSSB}. It is well known that if the last event time is right censored then the distribution of $T$ is not ``nonparametrically estimable'' on $[\tau_n, \infty)$.   Thus all finite observed times are in $[0,\tau_n]$ and we can only estimate the truncated version of $f(t|\bm x)$ on $[0,\tau_n]$,   $\bar f(t|\bm x)=f(t|T\in[0,\tau_n], \bm x)= {f(t|\bm x)}/{F(\tau_n|\bm x)}$, $t\in[0,\tau_n]$.  In many applications with right censored last observation $\bar f(t|\bm x)$ does not approximate $f(t|\bm x)$ because $F(\tau_n|\bm x)$ may be not close to one. 

\subsection{Approximate Bernstein Polynomial Model}

The full likelihood (\ref{eq: loglikelihood ell(gamma,p) for PH model}) cannot be maximized without specifying $S(t|\bm x_0)$ using a finite dimensional model.
Traditional method approximates $S(t|\bm x_0)$ by step-function and treats the jumps at observations as unknown parameters. For censored or other types of incompletely observed data this parametrization is difficult and complicated. However the Bernstein polynomial approximation  makes the parametrization simple and much easy \citep{Guan-jns-2015,Guan-2017-jns}.
Given any $\bm x_0$, 
we approximate the truncated density $\bar f(t|\bm x_0)=f(t|\bm x_0)/F(\tau_n|\bm x_0)$ by
$\bar f_m(t|\bm x_0;\bar{\bm p})=\tau_n^{-1}\sum_{i=0}^m \bar p_i\beta_{mi}(t/\tau_n)$, a mixture  of  beta densities $\beta_{mi}$ with shape parameters $(i+1, m-i+1)$, $i\in \mathbb{I}_0^m$, and unknown mixing proportions $\bar{\bm p}=\bar{\bm p}(\bm x_0)= (\bar p_0,\ldots,\bar p_m)$.  Here the dependence of $\bar{\bm p}=\bar{\bm p}(\bm x_0)$ on $\bm x_0$ will be suppressed.
The mixing proportions  $\bar{\bm p}$
 are subject to constraints   $\bar{\bm p}\in \mathbb{S}_m\equiv \{(u_0,\ldots,u_m)^\T\in \mathbb{R}^{m+1}: u_i\ge 0, \sum_{i=0}^m u_i=1.\}.$
Denote $\pi=\pi(\bm x_0)=F(\tau_n|\bm x_0)$.
Reparametrizing with $p_i=\pi\bar p_i$, $i\in \ints(0,m)$, we can approximate $f(t|\bm x_0)$   on $[0,\tau_n]$  by $f_m(t|\bm x_0;\bm p)=\pi(\bm x_0)\bar f_m(t|\bm x_0;\bm p)= \frac{1}{\tau_n}\sum_{i=0}^{m} p_i\beta_{mi}(t/\tau_n)$.
If $\pi< 1$,  although we do not need and cannot estimate the values of $f(t|\bm x_0)$  on $(\tau_n,\infty)$, we can put an arbitrary guess on them such as
$f_m(t|\bm x_0; \bm p)=  p_{m+1} \alpha(t-\tau_n)$, $t\in (\tau_n,\infty)$, where $p_{m+1}=1-\pi$ and $\alpha(\cdot)$ is a density on $[0,\infty)$ such that $(1-\pi)\alpha(0)=(m+1)p_m/\tau_n$ so that $f_m(t|\bm x_0; \bm p)$ is continuous at $t=\tau_n$, e.g., $\alpha(t)=\alpha(0)\exp[-\alpha(0)t]$. 
Thus $f(t|\bm x_0)$ and $S(t|\bm x_0)
$ on $[0,\infty)$, can be ``approximated'', respectively, by
\begin{equation}\label{eq: fm(t|x0; p) on (0,Inf)}
 f_m(t|\bm x_0;\bm p)
=\left\{
\begin{array}{ll}
   \frac{1}{\tau_n}\sum_{i=0}^{m} p_i\beta_{mi}(t/\tau_n), & \hbox{$t\in[0,\tau_n]$;} \\
p_{m+1} \alpha(t-\tau_n) , & \hbox{$t\in(\tau_n,\infty)$,}
\end{array}
\right.
\end{equation}
 and
\begin{equation}\label{eq: Sm(t|x0; p) on (0,Inf)}
S_m(t|\bm x_0;\bm p)
= \left\{
\begin{array}{ll}
  \sum_{i=0}^{m+1}  p_i\bar {\mathcal{B}}_{mi}(t/\tau_n), & \hbox{$t\in[0,\tau_n]$;} \\
p_{m+1} \bar {\mathcal{A}}(t-\tau_n) , & \hbox{$t\in(\tau_n,\infty)$.}
\end{array}
\right.
\end{equation}
where 
$\bar {\mathcal{B}}_{mi}(t) =1- {\mathcal{B}}_{mi}(t)=1-\int_0^t\beta_{mi}(s)ds$,  $i\in\ints(0,m)$, $\bar {\mathcal{B}}_{m,m+1}(t)\equiv 1$, and $\bar {\mathcal{A}}(t)=\int_t^\infty \alpha(u)du$.    
Thus we can approximate $S(t|\bm x)$ and $f(t|\bm x)$ on $[0,\tau_n]$, respectively, by
\begin{align}\label{generalized  proportional odds-rate model-approx survival function}
  S_m(t|\bm x; \bm\gamma, \bm p) &= S(t|\bm x; \bm\gamma, f_m(\cdot|\bm x_0;\bm p)), \\
\label{generalized  proportional odds-rate model-approx density}
 f_m(t|\bm x; \bm\gamma, \bm p) &=  f(t|\bm x; \bm\gamma, f_m(\cdot|\bm x_0;\bm p)).
\end{align}

If $\tau$ is finite and known we choose $\tau_n=\tau$ and specify $p_{m+1}=0$. Otherwise,  we choose $\tau_n=y_{(n)}$. In this case, from (\ref{eq: loglikelihood ell(gamma,p) for PH model}) we see that for data without right-censoring and covariate
we have to specify $p_{m+1}=0$ due to its unidentifiability.
If $\tau_n\ne 1$ we divide all the observed times by $\tau_n$.  Thus we assume  $\tau_n=1$ in the following.
We define $m^*=m$ or $=m+1$ according to whether we specify $p_{m+1}=0$ or not. Thus $\bm p=(p_0,\ldots,p_{m^*})$ and satisfies constraints
\begin{equation}\label{eq: constr for p=(p0,...,p_{m+1})}
\bm p=\bm p(\bm x_0)=(p_0,\ldots,p_{m^*})\in\mathbb{S}_{m^*}, \quad \mbox{$0\le p_{m+1}<1$}.
\end{equation}

The loglikelihood
$\ell(\bm\gamma, f_0;\bm z)$ can be approximated by
 the Bernstein loglikelihood $\ell_m(\bm\gamma, \bm p;\bm z)=\ell(\bm\gamma,  f_m(\cdot|\bm x_0; \bm p);\bm z)$, that is,
\begin{align*}
\ell_m(\bm\gamma,  \bm p;\bm z)&= ~ (1-\delta) [{\bm\gamma^\T \tilde{\bm x}} +\log f_m(y|\bm x_0; \bm p) +(e^{\bm\gamma^\T \tilde{\bm x}}-1)\log S_m(y|\bm x_0; \bm p)]\\\nonumber
&~~~ +  {\delta} \log[S_m(y_{1}|\bm x_0; \bm p)^{e^{\bm\gamma^\T \tilde{\bm x}}}-S_m(y_{2}|\bm x_0; \bm p)^{e^{\bm\gamma^\T \tilde{\bm x}}}],
\end{align*}
where $S_m(\infty|\bm x_0; \bm p)=0$.
 The  loglikelihood
$ \ell(\bm\gamma, f_0)= \sum_{i=1}^{n}\ell(\bm\gamma, f_0;\bm z_i)$
can be approximated by
$$ \ell_m(\bm\gamma, \bm p) = \sum_{i=1}^n\ell_m(\bm\gamma, \bm p;\bm z_i).$$
For a given degree $m$, if $(\hat{\bm\gamma}, \hat{\bm p})$ maximizes $\ell_m(\bm\gamma, \bm p)$ subject to constraints in (\ref{eq: constr for p=(p0,...,p_{m+1})}) for some ${\bm x}_0$ then $(\hat{\bm\gamma}, \hat{\bm p})$ is called the maximum approximate Bernstein (or beta) likelihood estimator (MABLE) of $(\bm\gamma, \bm p)$. This is a full likelihood method. The MABLE's of $f(t|\bm x)$ and $S(t|\bm x)$ are, respectively,
\begin{equation}\label{eq: f-hat and S-hat}
\hat  f_{\mathrm{B}}(t|\bm x)=f_m(t|\bm x; \hat {\bm\gamma}, \hat {\bm p}),\quad \hat  S_{\mathrm{B}}(t|\bm x)=S_m(t|\bm x; \hat{\bm\gamma}, \hat{\bm p}).
\end{equation}
The derivative of $\ell_m(\bm\gamma, \bm p;\bm z)$ with respect to $\bm p$ is
\begin{align}\nonumber\label{eq: deriv of ell(gamma,p) wrt p for GPO model}
\frac{\partial \ell_m(\bm\gamma, \bm p;\bm z)}{\partial\bm p} =&~ \bm\Psi(\bm\gamma, \bm p;\bm z)
\\
=&~(\Psi_0(\bm\gamma, \bm p;\bm z),\ldots,\Psi_{m^*}(\bm\gamma, \bm p;\bm z))^\T,
\end{align}
 where, 
for $j\in\ints(0,m^*)$,
\begin{align}\nonumber
\Psi_j(\bm\gamma,\bm p;\bm z) = &~
    {(1-\delta)}\Biggr[\frac{I(j\le m) {\beta}_{mj}(y)}{f_m(y|\bm x_0;\bm p)}+\frac{(e^{{\bm\gamma}^\T \tilde{\bm x}}-1) \bar {\mathcal{B}}_{mj}(y) }{S_m(y|\bm x_0; \bm p)}\Biggr]+ {\delta}
e^{{\bm\gamma}^\T \tilde{\bm x}} \\\label{eq: first m+1 term of gradient of ell}
&  \cdot \frac{S_m(y_{1}|\bm x_0; \bm p)^{e^{{\bm\gamma}^\T \tilde{\bm x}}\!-\!1}\bar {\mathcal{B}}_{mj}(y_{1})
-S_m(y_{2}|\bm x_0; \bm p)^{e^{{\bm\gamma}^\T \tilde{\bm x}}\!-\!1}\bar {\mathcal{B}}_{mj}(y_{2})}{S_m(y_{1}|\bm x_0; \bm p)^{e^{{\bm\gamma}^\T \tilde{\bm x}}}-S_m(y_{2}|\bm x_0;\bm p)^{e^{{\bm\gamma}^\T \tilde{\bm x}}}}.
\end{align}
\begin{lemma}\label{lem: Concavity of ell(p) for one obs}
The {Hessian matrix} $\bm H(\bm\gamma,\bm p)= \frac{\partial^2  \ell_m(\bm\gamma,  \bm p)}{\partial\bm p  \partial \bm p^\T} $ is nonpositive, i.e., all entries are nonpositive. 
For any fixed $\bm\gamma$ if ${\bm\gamma}^\T \bm x_0 \le \min_{1\le i\le n}\{{{\bm\gamma}^\T \bm x_i}\}$ then  $\bm H(\bm\gamma,\bm p)$ is negative semi-definite  for each $\bm p\in \mathbb{S}_{m^*}$. 
If, in addition, the vectors $[\Psi_j(\bm\gamma,\bm p;\bm z_1)$, $\ldots,\Psi_j(\bm\gamma,\bm p;\bm z_n)]$, $j\in\ints(0,m^*)$, are linearly independent, then $\bm H(\bm\gamma,\bm p)$ is negative definite.
\end{lemma}

Let  $\tilde{\bm p}=\tilde{\bm p}(\bm\gamma)=(\tilde p_0,\ldots,\tilde p_{m^*})^\T$ denote  the  maximizer of  $\ell_m(\bm\gamma, \bm p)$ with respect to $\bm p=(p_0,\ldots,p_{m^*})^\T$ subject to constraints in (\ref{eq: constr for p=(p0,...,p_{m+1})}).

Similar to \cite{Peters-and-Walker-1978-siam} we have the following result about a necessary and sufficient condition for $\tilde{\bm p}$.
\begin{theorem}\label{thm: necessary and sufficient condition}
For any fixed $\bm\gamma$ if ${\bm\gamma}^\T \bm x_0 \le \min_{1\le i\le n}\{{{\bm\gamma}^\T \bm x_i}\}$ then $\tilde{\bm p}=\tilde{\bm p}(\bm\gamma)$ is a maximizer of $ \ell_m(\bm\gamma, \bm p)$ if and only if
\begin{align}
\label{sufficient-necessary condition 2-interval censoring}
\lambda_n({\bm\gamma}):=\sum_{i=1}^ne^{{\bm\gamma}^\T \tilde{\bm x}_i}\ge &
\sum_{i=1}^n
\Psi_j ({\bm\gamma},  \tilde{\bm p};\bm z_i),
\end{align}
for all $j\in\mathbb{I}_0^{m^*}$ with equality if $\tilde p_j>0$.  If, in addition, the vectors $[\Psi_j(\bm\gamma,\bm p;\bm z_1)$, $\ldots,\Psi_j(\bm\gamma,\bm p;\bm z_n)]$, $j\in\ints(0,m^*)$, are linearly independent for all $\bm p$ in the interior of $\mathbb{S}_{m^*}$, then $\tilde{\bm p}$ is unique.
\end{theorem}
So it is necessary that
$\tilde  p_j =\tilde p_j  \bar\Psi_j({\bm\gamma}, \tilde {\bm p})$, $j\in\mathbb{I}_0^{m^*}$,
where
\begin{align*}
\bar\Psi_j({\bm\gamma}, {\bm p}) = \frac{1}{\lambda_n({\bm\gamma})}\frac{\partial\ell_m}{\partial p_j}(\bm\gamma, {\bm p})  = & \frac{1}{\lambda_n({\bm\gamma})}\sum_{i=1}^n\Psi_j ({\bm\gamma},  {\bm p};\bm z_i).
\end{align*}
We have fixed-point iteration
\begin{align}
\label{eq: interation for PH model with general censoring}
    p_j^{[s+1]} &= p_j^{[s]}\bar\Psi_j({\bm\gamma}, \bm p^{[s]}),
    \quad\quad\Big. j\in\ints(0,{m^*}),\quad s\in\ints(0,\infty),
\end{align}
If $ {\bm\gamma}^\T \bm x_0 \le \min_{1\le i\le n}\{{{\bm\gamma}^\T \bm x_i}\}$ then $\bar\Psi_j({\bm\gamma}, {\bm p})\ge 0$ for all $j\in\ints(0,{m^*})$
and $\bm p\in \mathbb{S}_{m^*}$.

Similar to the proof of Theorem 4 of \cite{Peters-and-Walker-1978-siam} we can prove the convergence of ${\bm p}^{[s]}$.
\begin{theorem}\label{thm: convergence of the fixed-point iteration}
For any fixed $\bm\gamma$ suppose   ${\bm\gamma}^\T \bm x_0 \le \min_{1\le i\le n}\{{{\bm\gamma}^\T \bm x_i}\}$. If $\bm p^{[0]}$ is in the interior of $\mathbb{S}_{m^*}$, the sequence $\{\bm p^{[s]}\}$ of (\ref{eq: interation for PH model with general censoring}) converges to 
$\tilde{\bm p}$.
\end{theorem}
Define an empirical $\bm\gamma$-related ``baseline'' $\hat{\bm x}_0=\hat{\bm x}_0(\bm\gamma)$ such that ${\bm\gamma}^\T \hat{\bm x}_0 =\min_{1\le i\le n}\{{{\bm\gamma}^\T \bm x_i}\}$.

\begin{lemma}\label{lem: concavity of ell wrt gamma}
The matrix  $\frac{\partial^2\ell_m(\bm\gamma, \bm p)}{\partial\bm\gamma\partial\bm\gamma^\T}$ is negative definite.
\end{lemma}
Let $\tilde {\bm\gamma}$ be an efficient estimator  of $\bm\gamma$  such as the NPMLE and SPMLE. We choose ${\bm x}_0=\hat{\bm x}_0(\tilde{\bm\gamma})$.
Then we maximize $\ell_m(\tilde{\bm\gamma},\bm p)$ to obtain $\tilde{\bm p}=\tilde{\bm p}(\tilde{\bm\gamma})$.
Therefore we can estimate
$f(t|\bm x)$ and $S(t|\bm x)$ on $[0,1]$, respectively, by
\begin{align}\nonumber\label{eq: f-tilde}
\tilde  f_\mathrm{B}(t|\bm x)&=f_m(t|\bm x; \tilde {\bm\gamma},  \tilde {\bm p})\\
&= \exp(\tilde{\bm\gamma}^\T\tilde{\bm x})[S_m(t|\bm x_0; \tilde{\bm p})]^{\exp(\tilde{\bm\gamma}^\T\tilde{\bm x})-1} f_m(t;\tilde{\bm p}),\\\label{eq: S-tilde}
\tilde  S_\mathrm{B}(t|\bm x)&=S_m(t|\bm x; \tilde {\bm\gamma}, \tilde {\bm p})=[S_m(t|\bm x_0; \tilde{\bm p})]^{\exp(\tilde{\bm\gamma}^\T\tilde{\bm x})}.
\end{align}
For the data without covariate, we have $\hat{\bm \gamma}=\bm 0$. Then we have $\hat f_\B(t)=f_m(t|\bm x;\bm 0,\hat{\bm p})$ and $\hat S_\B(t)=S_m(t|\bm x;\bm 0,\hat{\bm p})$.

For the NPMLE or SPMLE $\tilde{\bm\gamma}$ of $\bm\gamma$, the profile estimates $(\tilde{\bm\gamma}, \tilde{\bm p})$ are close to  $(\hat{\bm\gamma}, \hat{\bm p})$ especially for large sample size.  Thus $(\tilde{\bm\gamma}, \tilde{\bm p})$ can be used as initial values to find $(\hat{\bm\gamma}, \hat{\bm p})$ by the following algorithm. Such procedure was also suggested by \cite{Huang-1996-aos}. 
\begin{itemize}
\item []
\begin{itemize}
  \item [Step 0:] Start with an initial guess 
$\bm\gamma^{(0)}$ of $\bm\gamma$.
Choose $\bm x_0^{(0)}=\hat{\bm x}_0(\bm\gamma^{(0)})$.   Use (\ref{eq: interation for PH model with general censoring}) with $\tilde{\bm\gamma}= \bm\gamma_{0}$, $\bm x_0=\bm x_0^{(0)}$, and  starting point $\bm p^{[0]}=\bm u_m\equiv (1,\ldots,1)/(m^*+1)$  to get $\bm p^{(0)}=\tilde{\bm p}$.
Set $s=0$
  \item [Step 1:] Find the maximizer $\bm\gamma^{(s+1)}$ of $\ell_m(\bm\gamma,\bm p^{(s)})$ using the Newton-Raphson method.
  \item [Step 2:] Choose $\bm x_0^{(s+1)}=\hat{\bm x}_0(\bm\gamma^{(s+1)})$ and $\tilde{\bm\gamma}= \bm\gamma^{(s+1)}$. If $\tilde{\bm\gamma}^\T\Delta \bm x_0 \equiv \tilde{\bm\gamma}^\T(\bm x_0^{({s+1})}-\bm x_0^{({s})})=0$ then $\bm p^{[0]}=\bm p^{(s)}$ otherwise  $p^{[0]}_i=C_mf_m(i/m|\bm x_0^{(s+1)};\tilde {\bm\gamma}, \bm p^{(s)})$, $i\in\ints(0,m)$, $p^{[0]}_{m+1}= (p^{(s)}_{m+1})^{e^{\tilde{\bm\gamma}^\T\Delta \bm x_0 }}$ if $m^*=m+1$,  where $C_m$ is chosen so that $\sum_{i=0}^mp^{[0]}_i=1-p^{[0]}_{m+1}$.  Then use (\ref{eq: interation for PH model with general censoring}) with  $\bm x_0=\bm x_0^{({s+1})}$ to get $\bm p^{(s+1)}=\tilde{\bm p}$.    If the so obtained $\bm p^{[0]}$ is not in the interior of $\mathbb{S}_{m^*}$ we set $\bm p^{[0]}= (\bm p^{[0]}+\epsilon \bm u_m)/(1+\epsilon)$ using a small $\epsilon>0$. Set $s=s+1$.
   \item [Step 3:] Repeat Steps 1 and 2 until convergence. The final $\bm\gamma^{(s)}$ and $\bm p^{(s)}$ are taken as the MABLE $(\hat{\bm\gamma}, \hat{\bm p})$ of $(\bm\gamma,\bm p)$ with baseline $\hat{\bm x}_0=\bm x_0^{(s)}$.
\end{itemize}
\end{itemize}
The concavities of $\ell_m(\bm\gamma, \bm p)$ with respect to $\bm\gamma$ and $\bm p$ ensure that the above iterative algorithm is a point-to-point map and the solution set contains single point.
Convergence  of $(\bm\gamma^{(s)},\bm p^{(s)})$ to $(\hat{\bm\gamma}, \hat{\bm p})$ is guaranteed by the Global Convergence Theorem \citep{Zangwill-1969-book-nonlin-prog}.

\subsubsection{Some Special Cases}
\textbf{Data Without Covariate:}\label{sect: interval-censoring without covariate}
 For interval-censored data without covariate,  $\bm z_i=(\bm y_i,\delta_i)$,   $i\in \ints(1,n)$. The iteration (\ref{eq: interation for PH model with general censoring}) reduces to
\begin{equation}\label{eq: EM-iteration for p-hat}
p_j^{(s+1)}
=\frac{p_j^{(s)}}{n} \sum_{i=1}^{n}\Psi_{j}(\bm p^{(s)}; \bm z_{i}),\quad j\in\ints(0,{m^*}),
\end{equation}
where
$$\Psi_{j}(\bm p; \bm z)=
  \frac{(1-\delta) \beta_{mj}(y)}{f_m(y;\bm p)}+\delta
  \frac{\bar {\mathcal{B}}_{mj}(y_1)-\bar {\mathcal{B}}_{mj}(y_2)}{S_m(y_1;\bm p)-S_m(y_2;\bm p)},
\quad j\in\ints(0,{m^*}),$$ 
  $f_m(t;\bm p)=\sum_{j=0}^m p_i\beta_{mj}(t)$, and $S_m(t;\bm p)
=\sum_{j=0}^{m^*} p_j\bar {\mathcal{B}}_{mj}(t)$.
\\\\
\textbf{Two-Sample Data:}
%
When $\bm x=x$ is binary,  $x=1$ for cases and $x=0$ for controls, we have
a two-sample PH model which specifies  $S(t|1)=
[S(t|0)]^{\exp(\gamma)}$.
In this case,
usually $\gamma\ge 0$ so that $\Psi_j ({\bm\gamma}, {\bm p};\bm z)$ is always positive for each $j$.
In case $\gamma<0$  we switch case and control data.

\subsection{Model Selection}
The change-point method for model degree selection \citep{Guan-jns-2015} applies for finding an optimal degree $m$ for a given regression model.
%
Let $M=\{m_0,\ldots,m_k\}$, $m_i=m_0+i$, $i\in\ints(0,k)$. For each $i\in\ints(0,k)$, fit the data to obtain $(\hat{\bm\gamma},\hat{\bm p})$ and $\ell_i=  \ell_{m_i}(\hat{\bm\gamma},\hat{\bm p})$.
The optimal degree $m$ is the maximizer $\hat m$ of
$$R(m_i)=k\log\left(\frac{\ell_k-\ell_0}{k}\right)-i\log\left(\frac{\ell_i-\ell_0}{i}\right)-(k-i)\log\left(\frac{\ell_k-\ell_i}{k-i}\right),
\quad i\in\ints(1,k),$$
where $R(m_k)=0$. Alternatively, 
we can replace $\ell_i$ by $\ell_{m_i}(\tilde{\bm\gamma},\tilde{\bm p})$ where $\tilde{\bm p}=\tilde{\bm p}(\tilde{\bm\gamma})$ for a fixed efficient estimate $\tilde{\bm\gamma}$ for all $i$. The resulting optimal degree is denoted by $\tilde m$. Then using $m=\hat m$ or $m=\tilde m$ we obtain  $(\hat{\bm\gamma},\hat{\bm p})$.

\section{Asymptotic Results}
\subsection{Some Assumptions and Conditions}

The following assumptions are needed to develop asymptotic theory. 
\begin{assump}\label{A1}
The support $\mathcal{X}$ of covariate $\bm X$ is  {compact}  and for each $\bm x_0\in\mathcal{X}$, $\E(\tilde{\bm X}\tilde{\bm X}^\T)$ is positive definite, where $\tilde{\bm X}=\bm X-\bm x_0$.
\end{assump}
\begin{assump}\label{A2}
For each $\bm x_0\in\mathcal{X}$ and $\tau_n>0$, there exist  $f_m(t|\bm x_0; {\bm p}_0)$
 and $\rho>0$ such that,
 uniformly in $t\in[0,\tau_n]$,
\begin{equation}\label{approx for f(t|x0)}
\frac{f_m(t|\bm x_0; {\bm p}_0)-f(t|\bm x_0)}{f(t|\bm x_0)}=\mathcal{O}(m^{-\rho/2}),
\end{equation}
where ${\bm p}_0=(p_{01},\ldots,p_{0m},p_{0,m+1})^\T$,  $p_{0i}=\pi(\bm x_0)\bar p_{0i}$, $i\in \ints(0,m)$, $p_{0,m+1}=1-\pi(\bm x_0)=S(\tau_n|\bm x_0)$.
\end{assump}
For any $\bm\gamma$, the compactness of  $\cal{X}$ ensures the existence of $\bm x_{\gamma}\in\arg\min\{\bm\gamma^\T\bm x: \bm x\in \cal{X}\}$. Boundedness of $\bm X$ is assumed in the literature, e.g. (A3)(b) of \cite{Huang-and-Wellner-1997}. The positive finiteness of $\E(\tilde{\bm X}\tilde{\bm X}^\T)$ assures the identifiability of $\bm\gamma$.

 Let $\mathcal{C}^{(r)}[0,1]$ be the class of functions which have $r$th continuous derivative $f^{(r)}$ on $[0,1]$. A function $f$ is said to be {\em $\alpha$--H\"{o}lder
continuous} with $\alpha\in(0,1]$ if $|f(x)-f(y)|\le C|x-y|^\alpha$ for some constant $C>0$. We have the following result.
\begin{lemma}\label{lemma: a sufficient existence condition for Bernstein model}
Suppose that $\varphi(t)=t^{a}(1-t)^{b}\varphi_0(t)$ is a density on $[0,1]$, $a$ and $b$ are nonnegative integers,  $\varphi_0\in \mathcal{C}^{(r)}[0,1]$, $r\ge 0$,  $\varphi_0(t)\ge b_0>0$, and $\varphi_0^{(r)}$ is $\alpha$-H\"{o}lder
continuous with $\alpha\in(0,1]$.
Then  there exists $\bm p_0\in \mathbb{S}_m$ such that uniformly in $t\in[0,1]$,  with $\rho=r+\alpha$,
\begin{equation}\label{Assumption A1 for f on [0,1]}
\frac{f_m(t;\bm p_0)-\varphi(t)}{\varphi(t)}=\mathcal{O}(m^{-\rho/2}).
\end{equation}
\end{lemma}
This lemma was proved in \cite{Wang-and-Guan-2019}.
This is a generalization of the result of \cite{Lorentz-1963-Math-Annalen} which requires a positive lower bound for $\varphi$, \ie, $a=b=0$.

If $\varphi(t)=\tau_n\bar f(\tau_n t|\bm x_0)=\tau_n f(\tau_n t|\bm x_0)/\pi(\bm x_0)$ as a density on $[0,1]$ fulfills the condition of Lemma \ref{lemma: a sufficient existence condition for Bernstein model}, then
 assumption \ref{A2} is fulfilled.
The condition of Lemma \ref{lemma: a sufficient existence condition for Bernstein model} seems only sufficient for \ref{A2}.

 In the following, all expectations $\E(\cdot)$ are taken with respect to the (joint) distribution of random variable(s) in upper case.
The following are the conditions for cases considered in the asymptotic results.
\setcounter{cond}{-1}
\begin{cond}\label{C0}
The event time $T$ is uncensored and  $\tau_n=\tau<\infty$.
\end{cond}
\begin{cond}\label{C1}
The event time $T$ is subject to Case 1 interval censoring.
Given $\bm X=\bm x$ the inspection time $Y$ has cdf $G_1(\cdot|\bm x)$ on $[\tau_l,\tau_u]$, $0<\tau_l<\tau_u=\tau_n<\tau\le\infty$, and
$$\E[O(Y|\bm X)]=\int_{\cal{X}}\int_0^\infty  O(y|\bm x)  dG_1(y |\bm x)dH(\bm x)<\infty.$$
\end{cond}
\begin{cond}\label{C2}
The event time $T$ is subject to Case $k$ ($k\ge 2$) interval censoring
Given $\bm X=\bm x$ the {\bf observed}  inspection times  $\bm Y=(Y_1, Y_2)$ have joint cdf $G_2(\cdot,\cdot|\bm x)$ on $\{\bm y=(y_1, y_2): 0<\tau_l\le y_1\le  y_2\le \tau_u\}$, $\tau_n=\tau_u<\tau$, and
$$\E[O(Y_1|\bm X)S(Y_1|\bm X)]=\int_{\cal{X}}\int_0^\infty  O(y_1|\bm x) S(y_1|\bm x) dG_{21}(y_1 |\bm x)dH(\bm x)<\infty,$$
where $G_{21}$ is the marginal cdf of $Y_1$.
\end{cond}
The condition about the support of the inspection times are similar to those of \cite{Huang-and-Wellner-1997}.
The next theorem is about the identifiability of the approximate model.
\begin{theorem}\label{thm: model identifiability}
Suppose that $\bm X$ is almost surely linearly independent on $\mathcal{X}$. Then for uncensored data both $\bm\gamma$ and $\bm p$ are identifiable. For censored data, if, in addition, the inspection time is continuous then both $\bm\gamma$ and $\bm p$ are identifiable.
\end{theorem}

\subsection{Some Statistical Distances}
Under condition \ref{C0}, define statistical distances
\begin{align*}
\chi^2_0(\bm p;\bm x_0)&= \E\Big\{\Big[\frac{f_m(T|\bm x_0; {\bm p})}{f(T|\bm x_0)}-1\Big]^2\Big\}= \int_0^\tau\Big[ \frac{ f_m(y|\bm x_0;\bm p)}{f(y|\bm x_0)}-1\Big]^2 f(y)dy,\\
D_{0j}^2(\bm\gamma, \bm p; \bm x_0)&= \E\Big\{|e^{{\bm\gamma}^\T \tilde{\bm X}}-1|^j\Big[\frac{S_m(T|\bm x_0; {\bm p})}{S(T|\bm x_0)}-1\Big]^2 \Big\},\quad j=0,1, \\
    D_0^2(\bm p;\bm x_0)&=\chi^2_0(\bm p;\bm x_0)+D_{01}^2(\bm\gamma_0, \bm p; \bm x_0),
\end{align*}
where $\bm\gamma_0$  is the true value of $\bm\gamma$.

Under condition \ref{C1},  we define a weighted version of the \cite{Anderson-and-Darling-1954-jasa} distance as
\begin{align*}
D_1^2(\bm p; \bm x_0)&= \E\left\{ \Big[\frac{S_m(Y|\bm x_0; \bm p)}{S(Y|\bm x_0)}-1\Big]^2O(Y|\bm X)\right\}.
  \end{align*}
Under condition \ref{C2}, we
define
  \begin{align*}
D_{21}^2(\bm p;\bm x_0)&= \E\left\{ \Big[\frac{S_m(Y_1|\bm x_0; \bm p)}{S(Y_1|\bm x_0)}-1\Big]^2O(Y_1|\bm X)S(Y_1|\bm X) \right\}, \\
D_{22}^2(\bm p;\bm x_0)&= \E\left\{ \Big[\frac{S_m(Y_2|\bm x_0; \bm p)}{S(Y_2|\bm x_0)}-1\Big]^2 S(Y_2|\bm X) \right\}, \\
D_2^2(\bm p;\bm x_0)&=\max\{D_{2i}^2(\bm p;\bm x_0): i=1,2\}.
  \end{align*}
In the following the same symbols  $C$ and  $C'$  may represent different constants in different places.

\begin{theorem}\label{thm: rate of convergence of mable (gamma-hat, p-hat)}
  Let $(\hat{\bm\gamma},\hat{\bm p})$ be the MABLE of $(\bm\gamma,\bm p)$ with degree $m\ge C n^{1/\rho}$ for some constant $C>0$.  Suppose that assumptions \ref{A1} and \ref{A2} are satisfied. For each $i=0,1,2$, and any $\epsilon\in(0,1/2)$,  under condition (C\,$i$), we have
  $\Vert\hat{\bm\gamma}-\bm\gamma_0\Vert^2
\le C n^{-1+\epsilon}$, a.s.  and
$D^2_i(\hat{\bm p};\hat{\bm x}_0)\le C n^{-1+\epsilon}$, a.s..
\end{theorem}
\begin{theorem}\label{thm: Asym normality of gamma-tilde}
Suppose that assumptions \ref{A1} and \ref{A2} are satisfied.   Let $\tilde{\bm\gamma}=\tilde{\bm\gamma}({\bm p}_0)$ be the maximizer of $\ell_m(\bm\gamma,\bm p_0)$ for some $\bm p_0$ that satisfies \ref{A2}.
For each $i=0,1,2$, under condition (C\,$i$), $\sqrt{n}(\tilde{\bm\gamma}-\bm\gamma_0)$ converges in distribution to $N(\bm 0,\mathcal{I}^{-1})$ as $n\to\infty$, where $\bm x_0\in\arg\min_{\bm x\in\mathcal{X}}\bm\gamma_0^\T \bm x$,  $\mathcal{I} =\E(\tilde{\bm X}\tilde{\bm X}^\T)$ under condition \ref{C0}; $\mathcal{I}=\E \{ [O(Y|\bm X)\Lambda^2(Y|\bm X) ]\tilde{\bm X}\tilde{\bm X}^\T  \}$ under condition \ref{C1}; and
\begin{align*}\label{eq: information for case k>=2}
\mathcal{I} &=\E\Big[\frac{\bm \Lambda^\T \bm M \bm \Lambda}{S(Y_1|\bm X)-S(Y_2|\bm X)} \tilde{\bm X}\tilde{\bm X}^\T \Big]\\
&\ge \E\Big\{\Big[\frac{ S^2(Y_1|\bm X)\Lambda^2(Y_1|\bm X)}{1-S(Y_1|\bm X)}+ \frac{S^2(Y_2|\bm X)\Lambda^2(Y_2|\bm X)}{S(Y_2|\bm X)} \Big]\tilde{\bm X}\tilde{\bm X}^\T \Big\}\\\newtag
&= \E\Big\{\Big[O(Y_1|\bm X)S(Y_1|\bm X)\Lambda^2(Y_1|\bm X) +  S(Y_2|\bm X)\Lambda^2(Y_2|\bm X) \Big]\tilde{\bm X}\tilde{\bm X}^\T \Big\}
\end{align*}
under condition \ref{C2},
where $$\bm \Lambda=\left(
             \begin{array}{c}
               \Lambda (Y_1|\bm X) \\
               \Lambda (Y_2|\bm X) \\
             \end{array}
           \right),\quad \bm M=\left(
             \begin{array}{cc}
               \frac{F(Y_2|\bm X)}{F(Y_1|\bm X)} S^2(Y_1|\bm X)& -S(Y_1|\bm X)S(Y_2|\bm X) \\
                 -S(Y_1|\bm X)S(Y_2|\bm X)&S(Y_1|\bm X)S(Y_2|\bm X) \\
             \end{array}
           \right)$$
\end{theorem}
\begin{rem}
For Cox's maximum partial likelihood estimator $\hat{\bm\gamma}_{cox}$ from uncensored data, the information is 
\begin{align*}
\mathcal{I}_{cox}&=\E(\bm X\bm X^\T)-\int  \frac{1}{\int_{\mathcal{X}} f(t|\bm x)dH(\bm x)} \Big[\int_{\mathcal{X}}\bm x f(t|\bm x)dH(\bm x)\Big]^{\otimes 2}dt\\
&=\E(\bm X\bm X^\T)-\int \frac{1}{f(t)} \Big[\int_{\mathcal{X}}\bm x f(t|\bm x)dH(\bm x)\Big]^{\otimes 2}dt\\
&=\E[\Var(\bm X|T)]. 
\end{align*}
By the law of total covariance
$$\mathcal{I}_{cox}\le \E[\Var(\bm X|T)]+\Var[\E(\bm X|T)]=\Var(\bm X) \le \E(\tilde{\bm X}\tilde{\bm X}^\T)$$ 
with equality iff $\E(\bm X|T=t)$ is constant.
So under this surreal situation, the information $\mathcal{I}=\E(\tilde{\bm X}^{\otimes 2})\ge \mathcal{I}_{cox}$ for all $\bm x_0\in \mathcal{X}$. More  theoretical work need be done to access the information loss due to the unknown $\bm p_0$.
\end{rem}
Because $\ell_m(\bm\gamma,\bm p)$ depends on $\bm p$ through $f_m(\cdot|\bm x_0;\bm p)$ and $f_m(\cdot|\bm x_0;\bm p_0)\approx f_m(\cdot|\bm x_0;\hat{\bm p})$, although $\bm p_0$ is unknown, 
we have $\hat{\bm\gamma}\approx \tilde{\bm\gamma}$.
We can estimate the information $\mathcal{I}$ by, with $\bm x_0=\hat {\bm x}_0$,
$$\hat{\mathcal{I}}=\frac{1}{n}\sum_{i=1}^n\frac{\partial^2\ell_m(\hat{\bm\gamma}, \hat{\bm p};\bm z_i)}{\partial\bm\gamma\partial\bm\gamma^\T}.$$

\section{Simulation}\label{sect: Simulation}

Assume that given $\bm X=\bm x$, $T$ is Weibull $W(\theta, \sigma e^{-\bm\gamma^\T \bm x/\theta})$ so that the baseline $\bm x=\bm 0$ distribution is $W(\theta, \sigma)$ with shape and scale $\theta=\sigma=2$.
The function \verb"simIC_weib()" of R package \verb"icenReg" \citep{Anderson-Bergman-2017-JSS} 
 was used to generate interval censored data of sizes  $n=30, 50, 100$ with censoring probability is 70\%
from Weibull distributions. For data with covariate, $\bm X=(X_1,X_2)$, where $X_1$ and $X_2$ are independent, $X_1$ is uniform [-1,1] and $X_2=\pm 1$ is uniform,  with coefficients $\gamma_1=0.5$, $\gamma_2=-0.5$.  For data without covariate, \cite{Braun-etal-2005-cjs}'s kernel density estimation  implemented in R \verb"ICE" package was used.
In each case, 1000 samples were generated and used to estimate $\bm\gamma$, $f(\cdot|\bm 0)$ and $S(\cdot|\bm 0)$ on $[0, 7]$. If $\tau_n=y_{(n)}<7$ we use exponential $\alpha(\cdot)$ on $(\tau_n, 7)$ as in (\ref{eq: fm(t|x0; p) on (0,Inf)}) and (\ref{eq: Sm(t|x0; p) on (0,Inf)}).

\begin{figure}
\centering
  \makebox{\includegraphics[width=4.7in]{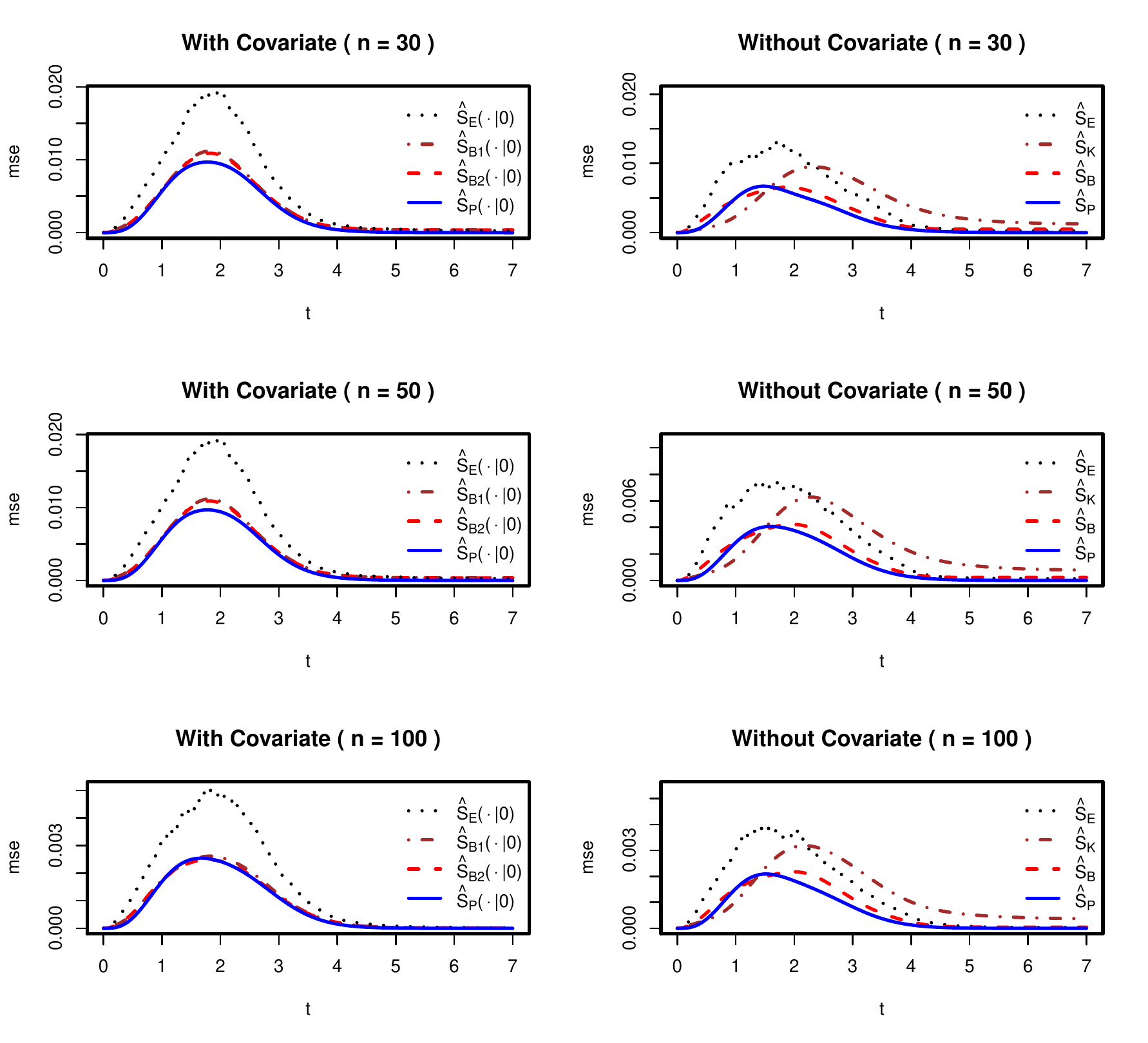}}
\caption{Simulated pointwise mean squared errors. Left panels: MSE of estimates of survival function at baseline ($\bm x=\bm 0$) using the MABLEs $\hat {S}_{\mathrm{B}_1}$ with $m=\tilde m$,  $\hat S_{\mathrm{B}_2}$ with $m=\hat m$, SPMLE $\hat{S}_{\mathrm{E}}$,  and the PMLE $\hat {S}_{\mathrm{P}}$.
Right panels: MSE of estimates of survival function without covariate using NPMLE $\hat{S}_{\mathrm{E}}$,  the MABLE $\hat {S}_{\mathrm{B}}$ with $m=\hat m$, the kernel estimate $\hat{S}_{\mathrm{K}}$,  and the PMLE $\hat {S}_{\mathrm{P}}$.}\label{fig: simulated mse of survival function of weibull data}
\end{figure}

The simulation results on the estimation of the regression coefficients are summarized in Table \ref{tbl: rMSE of estimates of gamma}. The pointwise mean squared errors of the estimated survival functions are plotted
in Figure \ref{fig: simulated mse of survival function of weibull data}. Since the proposed $\hat S_{\mathrm{B}}$ has smaller variance than the discrete SPMLE especially when sample size is not large,  the new estimator $\hat{\bm\gamma}$ may have smaller standard deviation than the traditional one. This is convinced by the simulation.
 From these results we see that the proposed estimates are better than the semiparametric estimates of $\gamma$'s and are close to the parametric maximum likelihood estimates(PMLEs) especially for small sample data. The two proposed estimates using $m=\tilde m$ and $m=\hat m$ are very close. The proposed method is compared with the kernel smoothing method of \cite{Braun-etal-2005-cjs} (see the right panels of Figure \ref{fig: simulated mse of survival function of weibull data}). The overall performance of the proposed method is close, and getting closer as sample size increases, to the PMLE and much better than the NPMLE and the kernel estimates.

\begin{table}
  \centering
  \caption{Mean squared errors of estimates of the regression coefficients using semiparametric method (SP), the proposed method using $m=\tilde m$ (B1), the proposed method using $m=\hat m$ (B2), and the parametric method (P).}\label{tbl: rMSE of estimates of gamma}
  \begin{tabular}{lccccccc}
    \hline
 &\multicolumn{3}{c}{$\gamma_1$} &&\multicolumn{3}{c}{$\gamma_2$}\\\cline{2-4}\cline{6-8}
Method &$n=30$ &$n=50$ &$n=100$ &&$n=30$ &$n=50$ &$n=100$\\
SP & 0.2799 & 0.1202 & 0.0467 & & 0.1038 & 0.0478 & 0.0184\\
B1 & 0.2392 & 0.1095 & 0.0469 & & 0.0883 & 0.0443 & 0.0175\\
B2 & 0.2380 & 0.1090 & 0.0461 & & 0.0868 & 0.0439 & 0.0174\\
P & 0.2184 & 0.0973 & 0.0437 & & 0.0756 & 0.0389 & 0.0163\\
    \hline
  \end{tabular}
\end{table}

\section{Examples}
\subsection{\cite{Gentleman-and-Geyer-1994-bka}'s Example}
\cite{Gentleman-and-Geyer-1994-bka} gave an artificial data set to show that Turnbull's nonparametric
 maximum likelihood estimator $\hat F(t)$ exists, but there are two fixed points of Turnbull's selfconsistency
 algorithm. The data consist of six intervals (0, 1), (0, 2), (0, 2), (1, 3), (1, 3), (2, 3).
Since there is no right-censored event time, $p_{m+1}=0$. Choosing $\tau_n=3$ we have the transformed intervals are $(y_{i1},y_{i2}):$
 $(0, 1/3), (0, 2/3), (0, 2/3), (1/3, 1), (1/3, 1), (2/3, 1).$
Let $q_1(\bm p)=\sum_{j=0}^m p_j \mathcal{B}_{mj}(1/3)$ and $q_2(\bm p)=\sum_{j=0}^m p_j \mathcal{B}_{mj}(2/3)$, where $\bm p=(p_0,p_1,\ldots,p_m)$.
The likelihood is
 $\ell_m(\bm p)=\ell(q_1,q_2)=\log q_1+2\log q_2+2\log(1-q_1)+\log(1-q_2).$
It attains maximum $-3.819085$ at $(q_1,q_2)=(1/3,2/3)$. So $\ell_m(\bm p)$ is maximized
whenever $q_1=\sum_{j=0}^m p_j \mathcal{B}_{mj}(1/3)=1/3$ and $q_2=\sum_{j=0}^m p_j \mathcal{B}_{mj}(2/3)=2/3$.
%
\begin{figure}[h]
\centering
 \includegraphics[width=5.0in]{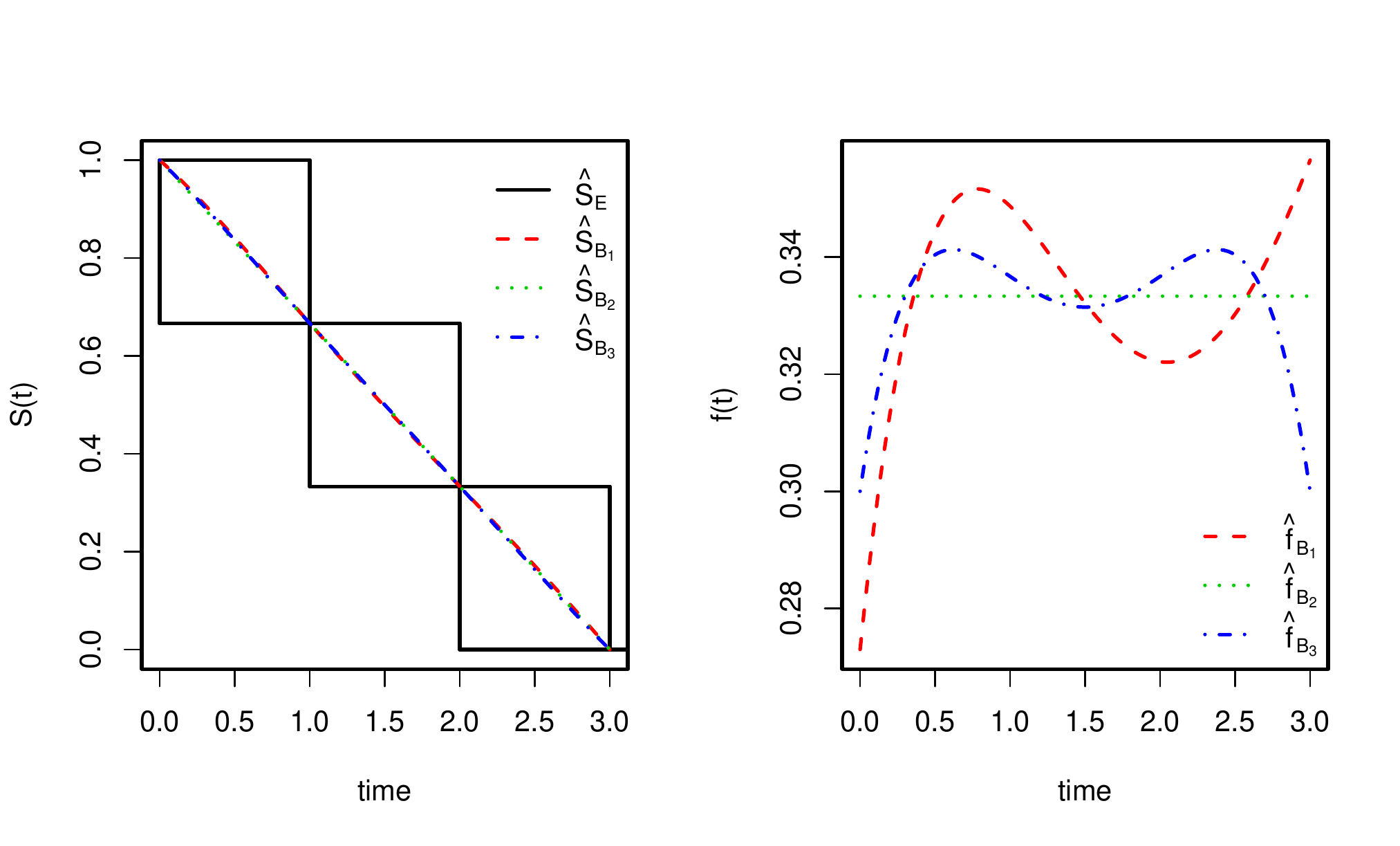}
\caption{\cite{Gentleman-and-Geyer-1994-bka}'s Example.
Left panel:  the NPMLE $\hat S_{\mathrm{E}}(\cdot|\bm x)$, the MABLEs $\hat S_{\mathrm{B}_i}$
with $m=6$ using initial $\bm p_i^{[0]}$, and
right panel:   the MABLEs $\hat f_{\mathrm{B}_i}$ using initial $\bm p_i^{[0]}$, $i=1,2,3$, where $\bm p_1^{[0]}=(1,2,\ldots,7)/28$,   $\bm p_2^{[0]}=(1,1,\ldots,1)/7$,  and $\bm p_3^{[0]}=(1,2,3,4,3,2,1)/16$.
}\label{fig: Gentleman-Geyer-Example}
\end{figure}
For this artificial dataset, the MABLE of $\bm  p$ is unique and uniform if $m=1,2$ but not unique if $m\ge 3$. Figure \ref{fig: Gentleman-Geyer-Example} shows the NPMLE of $S(t)$ and the MABLEs of $S(t)$ and $f(t)$ when $m=6$ with different starting points $\bm p_1^{[0]}=(1,2,\ldots,7)/28$,   $\bm p_2^{[0]}=(1,1,\ldots,1)/7$,  and $\bm p_3^{[0]}=(1,2,3,4,3,2,1)/16$.
Although the MABLE $\hat{\bm p}$ is not unique, as shown in Figure \ref{fig: Gentleman-Geyer-Example}, the resulting estimated survival functions are almost identical. A kernel density estimate for this dataset was discussed in \cite{Braun-etal-2005-cjs}.

\subsection{Stanford Heart Transplant Data}
To illustrate the use of the proposed method for right-censored data with binary covariate, we used the Stanford  Heart Transplant data  which is available in R \verb"survival" package. More information about this dataset can be found in \cite{Crowley-and-Hu-1977-jasa}.
%
%
%
We choose $X$, the indicator of prior bypass surgery,  as covariate and $\tau_n=y_{(n)}=1799$.  The Cox's partial likelihood estimate of $\gamma$ is $\tilde\gamma=-0.74072$ (s.e. 0.3591). With fixed $\gamma=\tilde\gamma$, the estimated degree is $\tilde  m=14$.  The MABLE  of $\bm p$ is $\tilde{\bm p}=(\tilde p_0,\ldots,\tilde p_{15})^\T$, where $\tilde p_0=0.470490$, $\tilde p_6=1.3256\times 10^{-6}$, $\tilde p_7=0.151148$, $\tilde p_8= 2.7997\times 10^{-5}$, $\tilde p_{10}=1.1001\times 10^{-7}$, $\tilde p_{11}=0.038977$, $\tilde p_{15}=1-\tilde \pi=0.339359$,  and all the other $\tilde p_i$'s are smaller than $10^{-9}$.
  Then we obtain
$$\tilde S_\mathrm{B}(t|\bm x=1)= S_{14}\left(\frac{t}{\tau_n}\Big|\bm x=1;\tilde{\bm p}\right)=\sum_{i=0}^{14}\tilde p_i \bar B_{{14},i}\left(\frac{t}{\tau_n}\right)+\tilde p_{15}.$$
\begin{figure}[ht]
\centering
\includegraphics[width=4.7in]{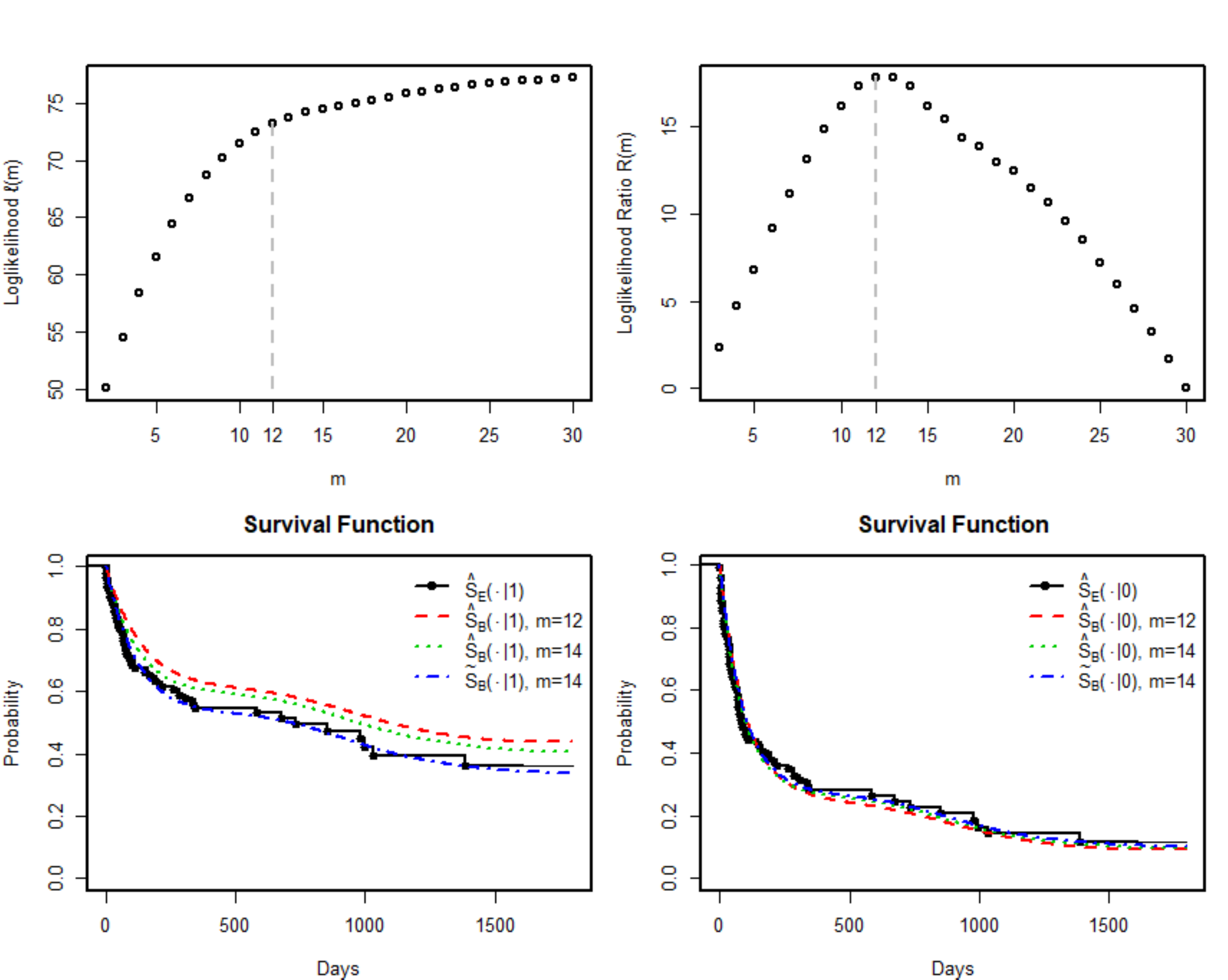}
\caption{Stanford heart transplant data.
Upper left panel: log-likelihood $\ell_m(\hat{\bm\gamma},\hat{\bm p})$; Upper right panel: likelihood ratio for choosing model degree using change-point estimate.
Lower panels: the SPMLE $\hat S_{\mathrm{E}}(\cdot|\bm x)$, the MABLEs $\hat S_{\mathrm{B}}(\cdot|\bm x)$ using $m=12$, $\hat S_{\mathrm{B}}(\cdot|\bm x)$ using $m=14$, and $\tilde S_{\mathrm{B}}(\cdot|\bm x)$ using $m=14$, with prior 
surgery  $\bm x=1$ (lower left) and without prior 
surgery  $\bm x=0$ (lower right).}\label{fig: stanford-heart-trans-data-Shat-Stilde-1-2}
\end{figure}
With the chosen $\tilde m=14$, the maximizer $(\hat{\bm\gamma}, \hat{\bm p})$ of $\ell_{\tilde m}(\bm\gamma, \bm p)$ was found to be $\hat\gamma=-0.95151$ (s.e. 0.12309) and $\hat{\bm p}=(\hat p_0,\ldots,\hat p_{15})^\T$, where $\hat p_0=0.40848$, $\hat p_2=4.49876\times 10^{-6}$, $\hat p_3=3.35856\times 10^{-6}$, $\hat p_6=1.12521\times 10^{-6}$, $\hat p_7=0.14646$, $\hat p_8= 2.28252\times 10^{-6}$, $\hat p_{10}=1.30873\times 10^{-6}$, $\hat p_{11}=0.03827$, $\hat p_{12}=1.21518\times 10^{-6}$, $\hat p_{15}=1-\tilde \pi=0.40677$,  and all the other $\hat p_i$'s are smaller than $10^{-6}$. The resulting estimated survival function is denoted by $\hat S_\mathrm{B}(t|\bm x=1)$ with $m=14$.

The optimal degree is $\hat m=12$ based on full likelihood $\ell_m(\hat{\bm\gamma}, \hat{\bm p})$.
The MABLE of $(\bm\gamma, \bm p)$ was found to be $\hat\gamma=-1.05959$ (s.e. 0.12309) and $\hat{\bm p}=(\hat p_0,\ldots,\hat p_{13})^\T$, where $\hat p_0=0.38968$,  $\hat p_6=0.11718$, $\hat p_7=0.02320$, $\hat p_8= 4.19865\times 10^{-6}$, $\hat p_{9}=0.03226$, $\hat p_{10}=5.74877\times 10^{-6}$,  $\hat p_{13}=1-\hat \pi=0.43767$,  and all the other $\hat p_i$'s are smaller than $10^{-6}$. The resulting estimated survival function is denoted by $\hat S_\mathrm{B}(t|\bm x=1)$ with $m=12$.
The results are shown in Figure  
\ref{fig: stanford-heart-trans-data-Shat-Stilde-1-2}. The proposed estimates of survival probabilities  for those who had (no) by-pass surgery are much larger (a little smaller) than the SPMLEs.

\subsection{Ovarian Cancer Data}
\begin{figure}
\centering
  \makebox{\includegraphics[width=4.7in]{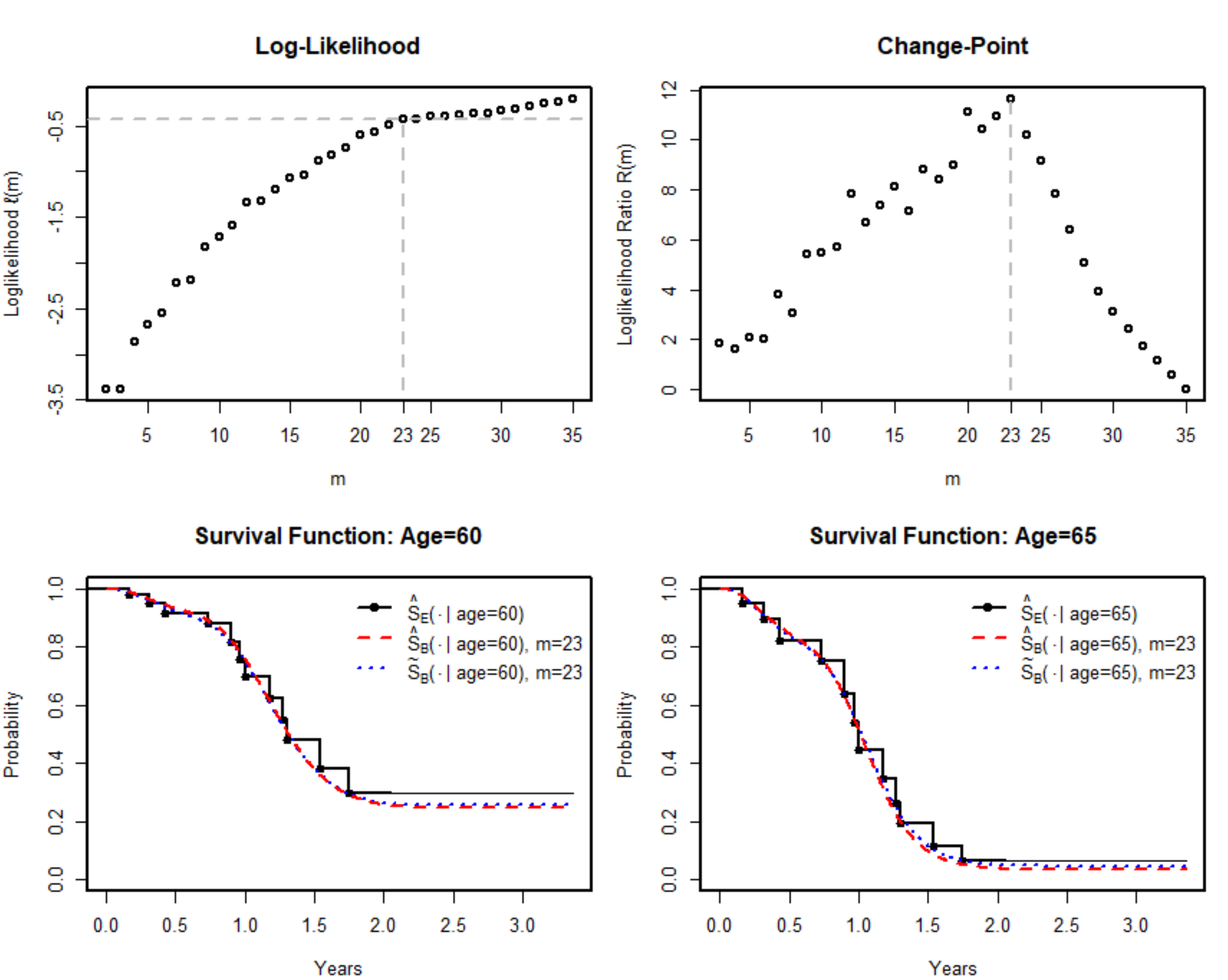}}
\caption{Ovarian cancer data.
Upper left panel: log-likelihood $\ell_m(\hat{\bm\gamma},\hat{\bm p})$; Upper right panel: likelihood ratio for choosing model degree using change-point estimate.
Lower panels: the SPMLE $\hat S_{\mathrm{E}}(\cdot|\bm x)$, the MABLEs $\hat S_{\mathrm{B}}(\cdot|\bm x)$,   and $\tilde S_{\mathrm{B}}(\cdot|\bm x)$, using $m=23$, given age  $\bm x=60$ (lower left) and  $\bm x=65$ (lower right).}\label{fig: ovarian-data}
\end{figure}

As an example of right-censored data with continuous covariate the ovarian cancer dataset contained in the R package \verb"Survival" \citep{survival-package} was originally reported by \cite{Edmonson1979DifferentCS}, and was used as real data example by several authors  \citep[e.g.][]{Collett-2003-book,Huang-and-Ghosh-2014-ssin}. 
 In this study $n = 26$ patients with advanced ovarian carcinoma (stages IIIB
and IV) were treated using either cyclophosphamide alone (1 g/m2) or
cyclophosphamide (500 mg/m2) plus adriamycin (40 mg/m2) by i.v. injection
every 3 weeks in order to compare the treatment effect in prolonging the time of survival. Twelve observations are uncensored and the rest is
right-censored. We choose $X$=Age. The Cox's partial likelihood estimate of $\gamma$ is $\tilde\gamma=0.16162$ (s.e.   0.04974). Using the proposed method we obtained optimal degree $m=23$ based on either $\ell_m(\tilde\gamma,\hat{\bm p})$ or $\ell_m(\hat\gamma,\hat{\bm p})$ (see upper panels of Figure \ref{fig: ovarian-data}). With $m=23$, we have $\hat\gamma=0.17665$ (
s.e. 0.01218), and $\hat x_0=38.89$. The components of $\hat{\bm p}$ are $\hat p_2=0.00226$,
$\hat p_9=0.02789$, $\hat p_{10}=0.00277$, $\hat p_{24}=0.96707$, and all the other $\hat p_i<10^{-6}$. The estimated survival curves given ages 60 and 65 are shown in Figure \ref{fig: ovarian-data}.

\section{Concluding Remarks}

We have seen that with a continuous approximate model it is much easy to write the full likelihood.
The parameter $\bm p$ is identifiable under some conditions. This overcomes the unidentifiability and roughness problem of the discrete NPMLE or SPMLE of survival function. Furthermore the proposed method gives better estimates of the regression coefficients. However, the discrete NPMLE or SPMLE is useful to obtain initial starting points for the proposed MABLEs of survival function and the regression coefficients.

\section{Appendix}
\subsection{Proof of Lemma \ref{lem: Concavity of ell(p) for one obs}}
Let $\bm p$ be any point in the interior of $\mathbb{S}_m$. For any {\em nonzero} vector $\bm v=(v_0,\ldots,v_{{m^*}})^\T\in \mathbb{R}^{m^*+1}$, define
$$w(y;\bm v)=\sum_{k=0}^m v_k\beta_{mk}(y),\quad W(y;\bm v)=\sum_{k=0}^{{m^*}} v_k\bar {\mathcal{B}}_{mk}(y).$$
By (\ref{eq: first m+1 term of gradient of ell}),  the $(j,k)$-entry of 
$\bm H(\bm\gamma,\bm p)$ is $H_{jk}=\sum_{i=1}^nH_{jk}(\bm z_i)$, where
\begin{align}\nonumber
H_{jk}(\bm z)&=\frac{\partial^2  \ell_m(\bm\gamma,\bm p;\bm z )}{\partial p_j\partial p_k} =
\frac{\partial\Psi_j(\bm\gamma, \bm p;\bm z)}{ \partial p_k}\\\nonumber
&= -{(1-\delta)}\Biggr[\frac{I(j,k\le m) {\beta}_{mj}(y){\beta}_{mk}(y)}{f_m^2(y|\bm x_0;\bm p)}+\frac{(e^{{\bm\gamma}^\T \tilde{\bm x}}-1) \bar {\mathcal{B}}_{mj}(y)\bar {\mathcal{B}}_{mk}(y) }{S_m^2(y|\bm x_0;\bm p)}\Biggr]\\\nonumber
& + {\delta}\Biggr\{e^{{\bm\gamma}^\T \tilde{\bm x}}  \Biggr[({e^{{\bm\gamma}^\T \tilde{\bm x}}-1})\\\nonumber
&\times \frac{S_m(y_{1};\bm p)^{e^{{\bm\gamma}^\T \tilde{\bm x}}-2}\bar {\mathcal{B}}_{mj}(y_{1})\bar {\mathcal{B}}_{mk}(y_{1})
-S_m(y_{2};\bm p)^{e^{{\bm\gamma}^\T \tilde{\bm x}}-2}\bar {\mathcal{B}}_{mj}(y_{2})\bar {\mathcal{B}}_{mk}(y_{2})}{S_m(y_{1}|\bm x_0;\bm p)^{e^{{\bm\gamma}^\T \tilde{\bm x}}}-S_m(y_{2}|\bm x_0;\bm p)^{e^{{\bm\gamma}^\T \tilde{\bm x}}}}\\\nonumber
&- e^{{\bm\gamma}^\T \tilde{\bm x}} \frac{S_m(y_{1}|\bm x_0;\bm p)^{e^{{\bm\gamma}^\T \tilde{\bm x}}-1}\bar {\mathcal{B}}_{mj}(y_{1})
-S_m(y_{2}|\bm x_0;\bm p)^{e^{{\bm\gamma}^\T \tilde{\bm x}}-1}\bar {\mathcal{B}}_{mj}(y_{2})}{S_m(y_{1}|\bm x_0;\bm p)^{e^{{\bm\gamma}^\T \tilde{\bm x}}}-S_m(y_{2}|\bm x_0;\bm p)^{e^{{\bm\gamma}^\T \tilde{\bm x}}}}\\\label{eq: (j,k)-entry of hessian of ell}
& \times \frac{S_m(y_{1}|\bm x_0;\bm p)^{e^{{\bm\gamma}^\T \tilde{\bm x}}-1}\bar {\mathcal{B}}_{mk}(y_{1})
-S_m(y_{2}|\bm x_0;\bm p)^{e^{{\bm\gamma}^\T \tilde{\bm x}}-1}\bar {\mathcal{B}}_{mk}(y_{2})}{S_m(y_{1}|\bm x_0;\bm p)^{e^{{\bm\gamma}^\T \tilde{\bm x}}}- S_m(y_{2}|\bm x_0;\bm p)^{e^{{\bm\gamma}^\T \tilde{\bm x}}}}\Biggr] \Biggr\}. 
\end{align}
Denote temporally $\eta=e^{{\bm\gamma}^\T \tilde{\bm x}}$, $B_{ij}=\bar {\mathcal{B}}_{mi}(y_{j};\bm v)$, and $V_j=S_m(y_{j}|\bm x_0; \bm p)$, $i\in\ints(0,{m^*})$,  $j=1,2$. We know $V_1\ge V_2$ and $B_{i1}\ge B_{i2}$.
In order to show that $H_{jk}(\bm z)\le 0$  for all $j,k\in\ints(0,{m^*})$, it suffices to show $A\le B$, where
$A=(V_1^{\eta-2}B_{j1}B_{k1}-V_2^{\eta-2}B_{j2}B_{k2})(V_1^\eta-V_2^\eta)$ and
$B=(V_1^{\eta-1}B_{j1}-V_2^{\eta-1}B_{j2})(V_1^{\eta-1}B_{k1}-V_2^{\eta-1}B_{k2})$. Now
\begin{align*}
    B-A
&\ge \frac{1}{2}V_1^{\eta-2}V_2^{\eta-2}[2V_2^2B_{j1}B_{k1}+2V_1^2B_{j2} B_{k2}\\
&~~~-(V_1^2+V_2^2)(B_{j1}B_{k2}+B_{j2}B_{k1})]\\
&\ge V_1^{\eta-2}V_2^{\eta}[B_{j1}B_{k1}+B_{j2}B_{k2} -(B_{j1}B_{k2}+B_{j2}B_{k1})] \\
&=V_1^{\eta-2}V_2^{\eta}(B_{j1}-B_{j2})(B_{k1}-B_{k2})\ge 0.
\end{align*}
For any $\bm v\in \mathbb{R}^{m^*+1}$, denoting $W_i=W(y_{i};\bm v)$, $i=1,2$, we have
$\bm v^\T \bm H(\bm\gamma,\bm p)\bm v=\sum_{i=1}^n \bm v^\T \bm H(\bm\gamma,\bm p;\bm z_i)\bm v$, where,
shown by simple algebra,
\begin{align*}
\bm v^\T \bm H(\bm\gamma,\bm p;\bm z)\bm v =&
-{(1-\delta)}\Biggr[\frac{w^2(y;\bm v)}{f_m^2(y|\bm x_0;\bm p)}+\frac{(\eta-1) W^2(y;\bm v)}{S_m^2(y|\bm x_0; \bm p)}\Biggr]\\\nonumber
& -{\delta}
\eta \Biggr[\frac{V_1^{\eta-2}W_1^2
-V_2^{\eta-2}W_2^2}{V_1^{\eta}-V_2^{\eta}}  + \eta  \frac{(V_2 W_1 - V_1 W_2)^2(V_1V_2)^{\eta-2}}{(V_1^{\eta}-V_2^{\eta})^2}  \Biggr].
\end{align*}
Since $\eta\ge 1$ we have
\begin{align*}
\bm v^\T \bm H(\bm\gamma,\bm p;\bm z)\bm v \le &
-{(1-\delta)}\Biggr[\frac{w^2(y;\bm v)}{f_m^2(y|\bm x_0;\bm p)}+\frac{(\eta-1) W^2(y;\bm v)}{S_m^2(y|\bm x_0; \bm p)}\Biggr]\\\nonumber
& -{\delta}
\eta  \Biggr(\frac{V_1^{\eta-1}W_1
-V_2^{\eta-1}W_2}{V_1^{\eta}-V_2^{\eta}}\Biggr)^2
\equiv -\bm v^\T \bm U_0(\bm\gamma,\bm p; \bm z)\bm v\le 0, 
\end{align*}
 where
$$\bm U_0(\bm\gamma,\bm p; \bm z)=(1-\delta)\Biggr[\frac{{\bm{\beta}_{m}(y)\choose 0}^{\otimes 2}}{f_m^2(y|\bm x_0;\bm p)}+\frac{(e^{{\bm\gamma}^\T \tilde{\bm x}}-1)\bm{ \bar {\mathcal{B}}}^{\otimes 2}_{m}(y) }{S_m^2(y|\bm x_0;\bm p)}\Biggr]+ \delta e^{{\bm\gamma}^\T \tilde{\bm x}} $$
$$ \times\left[\frac{S_m(y_{1}|\bm x_0;\bm p)^{e^{{\bm\gamma}^\T \tilde{\bm x}}-1}\bm{\bar {\mathcal{B}}}_{m}(y_{1})
-S_m(y_{2}|\bm x_0;\bm p)^{e^{{\bm\gamma}^\T \tilde{\bm x}}-1}\bm{\bar {\mathcal{B}}}_{m}(y_{2})}{S_m(y_{1}|\bm x_0;\bm p)^{e^{{\bm\gamma}^\T \tilde{\bm x}}}- S_m(y_{2}|\bm x_0;\bm p)^{e^{{\bm\gamma}^\T \tilde{\bm x}}}}\right]^{\otimes 2}.$$
Now $\bm v^\T \sum_{i=1}^n \bm U_0(\bm\gamma,\bm p; \bm z_i)\bm v=0$ implies,
for all $i\in\ints(1,n)$, $\sum_{j=0}^{m^*} v_j\Psi_{j}(\bm\gamma,\bm p;\bm z_i)=0$.
The proof of Lemma \ref{lem: Concavity of ell(p) for one obs} is complete.
\subsection{Proof of Lemma \ref{lem: concavity of ell wrt gamma}}
The derivatives of $\ell_m(\bm\gamma, \bm p;\bm z)$  with respect to $\bm\gamma$ are
\begin{align} \nonumber\label{eq: first derv of ell(gamma,p) for PH model}
 \frac{\partial\ell_m(\bm\gamma, \bm p;\bm z)}{\partial\bm\gamma}=&
  (1-\delta)[1+   e^{\bm\gamma^\T \tilde{\bm x}} \log S_m(y|\bm x_0;\bm p)]{\tilde{\bm x}}
  \\
  &+\delta
 \frac {\dot S_m(y_{1}|\bm x;\bm\gamma;\bm p)-\dot S_m(y_{2}|\bm x;\bm\gamma;\bm p)}{S_m(y_{1}|\bm x;\bm\gamma;\bm p)-S_m(y_{2}|\bm x;\bm\gamma;\bm p)},\\
\nonumber
 \frac{\partial^2\ell_m(\bm\gamma, \bm p;\bm z)}{\partial\bm\gamma\partial\bm\gamma^\T}=&
 (1-\delta) e^{\bm\gamma^\T \tilde{\bm x}} \log S_m(y|\bm x_0;\bm p){\tilde{\bm x}}\tilde{\bm x}^\T
\\\nonumber &+\delta \Big\{
\frac {\ddot S_m(y_{1}|\bm x;\bm\gamma;\bm p)-\ddot S_m(y_{2}|\bm x;\bm\gamma;\bm p)}{S_m(y_{1}|\bm x;\bm\gamma;\bm p)-S_m(y_{2}|\bm x;\bm\gamma;\bm p)}
 \\\label{eq: second derv of ell(gamma,p) for PH model}
  &- \frac {[\dot S_m(y_{1}|\bm x;\bm\gamma;\bm p)-\dot S_m(y_{2}|\bm x;\bm\gamma;\bm p)]^{\otimes 2}}{[S_m(y_{1}|\bm x;\bm\gamma;\bm p)-S_m(y_{2}|\bm x;\bm\gamma;\bm p)]^2} \Big\},
\end{align}
where
\begin{align}\label{eq: 1st derv of S(t|x) for PH model}
\dot S_m(t|\bm x; \bm\gamma;\bm p)
&= {e^{\bm\gamma^\T \tilde{\bm x}}} S_m(t|\bm x_0;\bm p)^{e^{\bm\gamma^\T \tilde{\bm x}}}\log S_m(t|\bm x_0;\bm p) \tilde{\bm x},\\\nonumber
\ddot S_m(t|\bm x; \bm\gamma;\bm p)
&= {e^{2\bm\gamma^\T \tilde{\bm x}}}
 S_m(t|\bm x_0;\bm p)^{e^{\bm\gamma^\T \tilde{\bm x}}}[\log S_m(t|\bm x_0;\bm p)]^2 \tilde{\bm x}\tilde{\bm x}^\T\\\label{eq: second derv of S(t|x) for PH model}
 &~~~+\dot S_m(t|\bm x; \bm\gamma;\bm p)\tilde{\bm x}^\T.
\end{align}
The lemma follows easily from (\ref{eq: second derv of ell(gamma,p) for PH model}) through(\ref{eq: second derv of S(t|x) for PH model}).
\subsection{Proof of Theorem \ref{thm: necessary and sufficient condition}}
If ${\bm\gamma}^\T \bm x_0 =\min_{1\le i\le n}\{{{\bm\gamma}^\T \bm x_i}\},$ we have ${\bm\gamma}^\T\tilde{\bm x}_i\ge 0$. By Lemma \ref{lem: Concavity of ell(p) for one obs}, $\ell_m(\bm\gamma, \bm p)$ is strictly concave on the compact and convex set $\mathbb{S}_{{m^*}}$ for the fixed $\bm\gamma$.
By the optimality condition for convex optimization \citep{Boyd-and-Vandenberghe-book-convex-optimization-2004}  we have that $\tilde{\bm p}$ is the unique maximizer of $\ell_m(\bm\gamma,\bm p)$ if and only if
\begin{equation}\label{sufficient-necessary condition 1-interval censoring}
\nabla_{\bm p}  \ell_m(\bm\gamma,\tilde{\bm p})^\T(\bm p-\tilde{\bm p})\le 0,\quad  \mbox{ for all $\bm p\in \mathbb{S}_{{m^*}}$},
\end{equation}
where $\nabla_{\bm p}  \ell_m(\bm\gamma, {\bm p})=\partial \ell_m(\bm\gamma,  {\bm p})/\partial \bm p$.
Therefore $\tilde{\bm p}$ is a maximizer of $\ell_m(\bm\gamma, {\bm p})$ for the fixed $\bm\gamma$ if and only if
\begin{align}
\label{sufficient-necessary condition-interval censoring}
\sum_{i=1}^ne^{{\bm\gamma}^\T \tilde{\bm x}_i}\ge & \frac{\partial \ell_m}{\partial p_j}(\bm\gamma, \tilde{\bm p})=
\sum_{i=1}^n
\Psi_j({\bm \gamma},\tilde{\bm p}; \bm z_i),
\end{align}
for all $j\in\mathbb{I}_0^{{m^*}}$ with equality if $\tilde p_j>0$.
The proof is complete.
\subsection{Proof of Theorem \ref{thm: convergence of the fixed-point iteration}}
Following the proof of Theorems 1 and 2 and the Corollary of \cite{Peters-and-Walker-1978-siam} we
define
 $\bm \Pi=\diag\{\bm p\}$ and
 ${\bm A}(\bm p,\bm\gamma)= \bm \Pi \nabla_{\bm p} \bar{\bm \Psi}(\bm p,\bm\gamma),$
where
$\bar{\bm \Psi}(\bm p,\bm\gamma)=[\bar \Psi_0(\bm p,\bm\gamma),\ldots, \bar \Psi_{{m^*}}(\bm p,\bm\gamma)]^\T$. Then
$${\bm A}(\bm p,\bm\gamma)=\frac{1}{\lambda_n(\bm\gamma)}\bm \Pi \nabla_{\bm p} \ell_m(\bm\gamma,\bm p).$$
Its gradient is
$$ \nabla {\bm A}(\bm p,\bm\gamma)=\frac{\partial {\bm A}(\bm p,\bm\gamma)}{\partial\bm p^\T}
=\frac{1}{\lambda_n(\bm\gamma)}\diag\{\nabla_{\bm p} \ell_m(\bm\gamma,\bm p)\}+\frac{1}{\lambda_n(\bm\gamma)}\bm\Pi\frac{\partial \nabla_{\bm p} \ell_m(\bm\gamma,\bm p)}{\partial\bm p^\T}
$$
$$=\frac{1}{\lambda_n(\bm\gamma)}\diag\Big\{\frac{\partial \ell_m(\bm\gamma,\bm p)}{\partial\bm p}\Big\}+\frac{1}{\lambda_n(\bm\gamma)}\bm\Pi\frac{\partial^2 \ell_m(\bm\gamma,\bm p)}{\partial\bm p\partial\bm p^\T}.$$
%
For any norm on $\mathbb{R}^{m^*+1}$ we have
$${\bm A}(\bm p,\bm\gamma)-\tilde {\bm p}=\nabla {\bm A}(\tilde {\bm p},\bm\gamma)(\bm p-\tilde {\bm p})
+\mathcal{O}(\Vert \bm p-\tilde {\bm p}\Vert^2). $$
Consider $\nabla {\bm A}(\tilde {\bm p},\bm\gamma)$ as an operator on subspace
$$\mathbb{Z}_m=\{\bm z\in R^{m^*+1} : \bm 1^\T\bm z=0\}. $$
If all components of $\tilde{\bm p}$ are positive then $\nabla_{\bm p} \ell_m(\bm\gamma,\tilde {\bm p})=\lambda_n(\bm\gamma)\bm 1$, and
 $\nabla {\bm A}(\tilde {\bm p},\bm\gamma)=I_{m^*+1}-\bm Q, $
where $$\bm Q=-\frac{1}{\lambda_n(\bm\gamma)}\tilde {\bm \Pi}\frac{\partial^2 \ell_m(\bm\gamma,\tilde {\bm p})}{\partial\bm p\partial\bm p^\T}.$$
From Lemma \ref{lem: Concavity of ell(p) for one obs} and (\ref{eq: (j,k)-entry of hessian of ell}) it follows that  $\bm Q$ is a left stochastic matrix and $\tilde {\bm p}^\T\frac{\partial^2 \ell_m(\bm\gamma,\tilde {\bm p})}{\partial\bm p\partial\bm p^\T}=
-\frac{\partial \ell_m(\bm\gamma,\tilde {\bm p})}{\partial\bm p^\T}=-\lambda_n(\bm\gamma)\bm 1^\T$. So
  $\mathbb{Z}_m$ is   invariant under  $\bm Q$.

Define an inner product $\langle\cdot,\cdot\rangle$ by $\langle \bm u,\bm v\rangle=\bm u^\T\tilde {\bm \Pi}^{-1}\bm v$ for $\bm u$, $\bm v$ in $\mathbb{Z}_m$. It can be easily shown that, with respect to this inner product, $\bm Q$ is symmetric and positive   semidefinite  on $\mathbb{Z}_m$:
$$\langle \bm u,\bm Q\bm v\rangle=\bm u^\T\tilde {\bm \Pi}^{-1}\bm Q\bm v=
-\frac{1}{\lambda_n(\bm\gamma)}\bm u^\T \frac{\partial^2 \ell_m(\bm\gamma,\tilde {\bm p})}{\partial\bm p\partial\bm p^\T}\bm v=\bm u^\T \bm Q^\T\tilde {\bm \Pi}^{-1}\bm v=\langle \bm Q\bm u,\bm v\rangle,
$$
$$\langle \bm u,\bm Q\bm u\rangle=
-\frac{1}{\lambda_n(\bm\gamma)}\bm u^\T \frac{\partial^2 \ell_m(\bm\gamma,\tilde {\bm p})}{\partial\bm p\partial\bm p^\T}\bm u
\ge 0.$$
Let $\mu_0$ and $\mu_m$ be the smallest and largest eigenvalues of $\bm Q$ associated with eigenvectors in $\mathbb{Z}_m$. Then the operator norm of $\nabla {\bm A}(\tilde {\bm p},\bm\gamma)$ on $\mathbb{Z}_m$ w.r.t. this inner product equals
$\max\{|1-\mu_0|,|1-\mu_m|\}$. It is clear that $0\le \mu_0\le\mu_m\le 1$ because $\bm Q$ is a left stochastic matrix. By Lemma \ref{lem: Concavity of ell(p) for one obs}  we have $\mu_0>0$.
Similar to the proof of Theorem 2 of \cite{Peters-and-Walker-1978-siam} the assertion of theorem follows. If $\tilde{\bm p}$ contains zero component(s),
say $\tilde p_j=0$, $j\in J_0$, deleting the $j$-th row  and $j$-th column  of the vectors and matrices in the above proof for all $j\in J_0$ we can show that  the iterates $p_j^{[s]}$, $s\in\ints(0,\infty)$, converge to $\tilde p_j$ as $s\to \infty$ for all $j\notin J_0$. Because $\sum_{j=0}^{m^*}  p_j^{[s]}=1$ and $ p_j^{[s]}\ge 0$, $j\in\ints(0,{m^*})$, for those $j\in J_0$,  $p_j^{[s]}$ converges to zero as $s\to \infty$.
The proof of Theorem \ref{thm: convergence of the fixed-point iteration} is complete.
\subsection{Proof of Theorem \ref{thm: model identifiability}}
 If $\ell_m(\bm\gamma^{(1)},  \bm p^{(1)};\bm z)\equiv \ell_m(\bm\gamma^{(2)},  \bm p^{(2)};\bm z)$, where $\bm\gamma^{(i)}\in\Gamma$ and $\bm p^{(i)}\in\mathbb{S}_{m^*}$, $i=1,2$, then (i) for uncensored data we have $f_m(y|\bm x_0; \bm p^{(1)})\equiv f_m(y|\bm x_0; \bm p^{(2)})$ and $\tilde{\bm x}^\T \bm\gamma^{(1)}\equiv\tilde{\bm x}^\T \bm\gamma^{(2)}$; and (ii) for censored data we have
$$S_m(y_{j}|\bm x_0; \bm p^{(1)})^{e^{\tilde{\bm x}^\T \bm\gamma^{(1)}}} \equiv S_m(y_{j}|\bm x_0; \bm p^{(2)})^{e^{\tilde{\bm x}^\T \bm\gamma^{(2)}}},\quad j=1,2.$$  For case (i) we have $\bm p^{(1)}=\bm p^{(2)}$ as shown by \cite{Guan-jns-2015} and $\bm\gamma^{(1)}=\bm\gamma^{(2)}$ if $\tilde{\bm x}$ is linearly independent. For case (ii) we have $S_m(y_{j}|\bm x_0; \bm p^{(1)}) \equiv S_m(y_{j}|\bm x_0; \bm p^{(2)})$ which implies $\bm p^{(1)}=\bm p^{(2)}$ and $\bm\gamma^{(1)}=\bm\gamma^{(2)}$ if $\tilde{\bm x}$ is linearly independent.

\subsection{Proof of Theorem \ref{thm: rate of convergence of mable (gamma-hat, p-hat)}}
We need the following lemma for the proof.
\begin{lemma}\label{lem: rate of convergence of gamma-tilde and fm(y|x0,p-tilde)}
Suppose that assumptions \ref{A1} and \ref{A2} with $m\ge C_0 n^{1/k}$ for some constant $C_0$, and condition (C$i$) are satisfied for an $i\in\ints(0,2)$ and an  $\epsilon\in(0,1/2)$. If $\Vert \bm\gamma-\bm\gamma_0\Vert^2\le C n^{-1+\epsilon}$  then for any $\epsilon'\in(\epsilon,1/2)$ and
$n$  large enough the maximizer $\tilde{\bm p}=\tilde{\bm p}(\bm\gamma)$ of $\ell_m(\bm\gamma, \bm p)$  almost surely satisfies $D_i^2(\tilde{\bm p}; \bm x_0)
\le C' n^{-1+\epsilon'}$, for some constant $C'>0$,
where $\bm x_0=\bm x_{\bm\gamma}$, $\tilde{\bm p}\in A_m(\epsilon_n)$.
Conversely,  if ${D_i^2(\tilde{\bm p}; \bm x_0)
\le C n^{-1+\epsilon}}$, for  some $\bm x_0$,  then  for any $\epsilon'\in(\epsilon,1/2)$ and
$n$  large enough the maximizer $\tilde{\bm\gamma}=\tilde{\bm\gamma}(\bm p)$ of
$\ell(\bm\gamma, \bm p)$  for the fixed $\bm p$ almost surely satisfies  $\Vert\tilde{\bm\gamma}-\bm \gamma_0\Vert^2\le C' n^{-1+\epsilon'}$, for some constant $C'>0$.
\end{lemma}

\subsubsection*{Proof of Lemma \ref{lem: rate of convergence of gamma-tilde and fm(y|x0,p-tilde)}}
Define
   $\ell(\bm\gamma,f_0)=\sum_{i=1}^n \ell(\gamma, f_0;\bm z_i)$ and
   ${\mathcal{R}}({\bm\gamma},\bm p)=\ell({\bm\gamma}_0,f_0)-\ell_m({\bm\gamma},\bm p)$.
By Taylor expansion we have, for all $\bm p\in \mathcal{A}(\epsilon_n)$,
  \begin{align*}\label{eq: log likelihood ratio}
   {\mathcal{R}}({\bm\gamma},\bm p)
     &=-\sum_{i:\delta_i=0}\Biggr[(\bm\gamma-\bm\gamma_0)^\T\tilde{\bm x}_i+ \log \frac{f_m(y_i|\bm x_0;\bm p)}{f(y_i|\bm x_0)} +(e^{{\bm\gamma}^\T \tilde{\bm x}_i}-1) \log \frac{S_m(y_i|\bm x_0;\bm p)}{ S(y_i|\bm x_0)} \\ &~~~ -(e^{{\bm\gamma}^\T \tilde{\bm x}_i}-e^{{\bm\gamma}_0^\T \tilde{\bm x}_i}) \Lambda(y_i|\bm x_0)\Biggr]\\ & ~~~ +\sum_{i:\delta_i=1} \Biggr[\log\frac{S_m(y_{1i}|\bm x_0;\bm p)^{e^{{\bm\gamma}^\T \tilde{\bm x}_i}}-S_m(y_{2i}|\bm x_0;\bm p)^{e^{{\bm\gamma}^\T \tilde{\bm x}_i}}}{ S(y_{1i}|\bm x_0)^{e^{{\bm\gamma}_0^\T \tilde{\bm x}_i}}-S(y_{2i}|\bm x_0)^{e^{{\bm\gamma}_0^\T \tilde{\bm x}_i}}}\Biggr]\\\newtag
&= \sum_{0\le i\le j\le 1}\tilde{\mathcal{R}}_{ij}({\bm\gamma},\bm p)+\frac{1}{2}\sum_{i=0}^1\tilde{\mathcal{R}}_{i2}({\bm\gamma},\bm p)+\sum_{i=0}^1o[\tilde{\mathcal{R}}_{i2}({\bm\gamma},\bm p)],
  \end{align*}
where
$\tilde{\mathcal{R}}_{00}({\bm\gamma},\bm p)=\sum_{i=1}^n (1-\delta_i) U_{0i}(\bm\gamma)$,
$\tilde{\mathcal{R}}_{01}({\bm\gamma},\bm p)=-\sum_{i=1}^n(1-\delta_i) [U_{1i}(\bm p) +(e^{{\bm\gamma}^\T \tilde{\bm x}_i}-1) U_{2i}(\bm p)]$,
$\tilde{\mathcal{R}}_{02}({\bm\gamma},\bm p)=\frac{1}{2}\sum_{i=1}^n(1-\delta_i) [U_{1i}^2(\bm p) +(e^{{\bm\gamma}^\T \tilde{\bm x}_i}-1) U_{2i}^2(\bm p)]$,
$\tilde{\mathcal{R}}_{11}({\bm\gamma},\bm p)=-\sum_{i=1}^n\delta_i U_{3i}({\bm\gamma},\bm p)$,
$\tilde{\mathcal{R}}_{12}({\bm\gamma},\bm p)=\frac{1}{2}\sum_{i=1}^n\delta_i U_{3i}^2({\bm\gamma},\bm p)$,
\begin{align*}
U_{0i}(\bm\gamma)&= (e^{{\bm\gamma}^\T \tilde{\bm x}_i}-e^{{\bm\gamma}_0^\T \tilde{\bm x}_i}) \Lambda(y_i|\bm x_0)-(\bm\gamma-\bm\gamma_0)^\T\tilde{\bm x}_i,\\
U_{1i}(\bm p)&= \frac{f_m(y_{i}|\bm x_0;\bm p)}{f(y_i|\bm x_0)}-1, \quad
U_{2i}(\bm p)=  \frac{S_m(y_{i}|\bm x_0;\bm p)}{S(y_i|\bm x_0)}-1,\\
U_{3i}(\bm\gamma,\bm p)&= \frac{S_m(y_{1i}|\bm x_0;\bm p)^{e^{{\bm\gamma}^\T \tilde{\bm x}_i}}-S_m(y_{2i}|\bm x_0;\bm p)^{e^{{\bm\gamma}^\T \tilde{\bm x}_i}}}{S(y_{1i}|\bm x_0)^{e^{{\bm\gamma}_0^\T \tilde{\bm x}_i}}-S(y_{2i}|\bm x_0)^{e^{{\bm\gamma}_0^\T \tilde{\bm x}_i}}}-1.
  \end{align*}
It is clear, for all real $x$,
\begin{equation}\label{eq: expansion of exp(x)-1}
\left| e^{x}-1
-\sum_{i=1}^j\frac{1}{i!}
x^i\right|\le \frac{e^{|x|}}{(j+1)!}\mathcal{O}(|x|^{j+1}),\quad j\in \ints(1,\infty).
\end{equation}
\paragraph{Proof of Lemma \ref{lem: rate of convergence of gamma-tilde and fm(y|x0,p-tilde)} under condition \ref{C0}:}

  For uncensored data, all $\delta_i=0$.
By integration by parts we have
  \begin{align*}\label{eq: E(U0)}
    \E[U_{0i}(\bm\gamma)]
&=\int_{\cal{X}}\left[({\bm\gamma}_0-{\bm\gamma})^\T \tilde{\bm x}- (e^{{\bm\gamma}_0^\T \tilde{\bm x}}-e^{{\bm\gamma}^\T \tilde{\bm x}})\int_0^\infty \log S(y|\bm x_0)  dS(y|\bm x) \right] dH(\bm x)\\
&=\int_{\cal{X}}\left[({\bm\gamma}_0-{\bm\gamma})^\T \tilde{\bm x}-   (e^{{\bm\gamma}_0^\T \tilde{\bm x}}-e^{{\bm\gamma}^\T \tilde{\bm x}})e^{-{\bm\gamma}_0^\T \tilde{\bm x}}\int_0^\infty f(y|\bm x)dy \right]dH(\bm x)\\
&=\int_{\cal{X}}\left[({\bm\gamma}_0-{\bm\gamma})^\T \tilde{\bm x}-   (e^{{\bm\gamma}_0^\T \tilde{\bm x}}-e^{{\bm\gamma}^\T \tilde{\bm x}})e^{-{\bm\gamma}_0^\T \tilde{\bm x}}  \right]dH(\bm x)\\\newtag
&=\int_{\cal{X}}  \left\{\sum_{i=2}^{j-1}\frac{[({\bm\gamma}-{\bm\gamma}_0)^\T \tilde{\bm x}]^i}{i!}
 +e^{|({\bm\gamma}\!-\!{\bm\gamma}_0)^\T \tilde{\bm x}|}\frac{\mathcal{O}[|({\bm\gamma}\!-\!{\bm\gamma}_0)^\T \tilde{\bm x}|^{j}]}{j!}\right\}dH(\bm x),
  \end{align*}
where $ j\in\ints(3,\infty)$.
Since $\bm X$ is bounded   we have, for all  $\bm\gamma\in \mathbb{B}_{d}(n^{-1+\epsilon})$,
\begin{equation}\label{ineq: estimates of E[U0i]}
\lambda_{0}\Vert{\bm\gamma}-{\bm\gamma}_0 \Vert^2\le \E[U_{0i}(\bm\gamma)]-o(\Vert{\bm\gamma}-{\bm\gamma}_0 \Vert^2)\le \lambda_{d}\Vert{\bm\gamma}-{\bm\gamma}_0 \Vert^2,
\end{equation}
where $\lambda_{0}>0$ and $\lambda_{d}>0$ are, respectively, the minimum  and maximum eigenvalues of $\E(\tilde{\bm X}\tilde{\bm X}^\T)$.
Similarly, repeated integration by parts implies
  \begin{align*}\label{eq: E(U0^2)}
    \E[U_{0i}^2(\bm\gamma)]&=\int_{\cal{X}}\left\{[({\bm\gamma}-{\bm\gamma}_0)^\T \tilde{\bm x}]^2  + (e^{{\bm\gamma}^\T \tilde{\bm x}}-e^{{\bm\gamma}_0^\T \tilde{\bm x}})^2\int_0^\infty  \Lambda^2(y|\bm x_0)   f(y|\bm x)dy\right\} dH(\bm x)\\
&~~~+2\int_{\cal{X}}  ({\bm\gamma}-{\bm\gamma}_0)^\T \tilde{\bm x}(e^{{\bm\gamma}^\T \tilde{\bm x}}-e^{{\bm\gamma}_0^\T \tilde{\bm x}}) \int_0^\infty  \Lambda(y|\bm x_0)   f(y|\bm x)dy dH(\bm x)\\
&=\int_{\cal{X}}[({\bm\gamma}-{\bm\gamma}_0)^\T \tilde{\bm x}]^2 dH(\bm x) +2\int_{\cal{X}}  (e^{{\bm\gamma}^\T \tilde{\bm x}-{\bm\gamma}_0^\T \tilde{\bm x}}-1)^2dH(\bm x)\\
&~~~-2\int_{\cal{X}}  ({\bm\gamma}-{\bm\gamma}_0)^\T \tilde{\bm x}(e^{{\bm\gamma}^\T \tilde{\bm x}-{\bm\gamma}_0^\T \tilde{\bm x}}-1)  dH(\bm x)\\
&=\int_{\cal{X}}\left\{\big[({\bm\gamma}-{\bm\gamma}_0)^\T \tilde{\bm x}-(e^{{\bm\gamma}^\T \tilde{\bm x}-{\bm\gamma}_0^\T \tilde{\bm x}}-1)\big]^2+ (e^{{\bm\gamma}^\T \tilde{\bm x}-{\bm\gamma}_0^\T \tilde{\bm x}}-1)^2\right\}dH(\bm x).
  \end{align*}
By (\ref{eq: expansion of exp(x)-1}) we have $|e^x-1-x|\le \frac{1}{2}|x|^2e^{|x|}$, and
\begin{equation}\label{ineq: upper bound for var(U0)}
\Var[U_{0i}(\bm\gamma)]\le \int_{\cal{X}}\Big[\frac{1}{4}|({\bm\gamma}-{\bm\gamma}_0)^\T \tilde{\bm x}|^4 e^{2|{\bm\gamma}^\T \tilde{\bm x}-{\bm\gamma}_0^\T \tilde{\bm x}|}+ (e^{{\bm\gamma}^\T \tilde{\bm x}-{\bm\gamma}_0^\T \tilde{\bm x}}-1)^2\Big]dH(\bm x).
\end{equation}
Consequently
\begin{equation}\label{ineq: the order of upper bound for var(U0)}
\Var[U_{0i}(\bm\gamma)]\le \eta' \lambda_{d} n^{-1+\epsilon}.
\end{equation}
Therefore
by LIL we have, for all $\bm\gamma\in\mathbb{B}_{d}(n^{-1+\epsilon})$,
  \begin{align*}\label{eq: estimate of R00}
\tilde{\mathcal{R}}_{00}({\bm\gamma},\bm p)&=\sum_{i=1}^nU_{0i}(\bm\gamma)\\
&=n\E[U_{0i}(\bm\gamma)]+\mathcal{O}(\sqrt{n\sigma^2[U_{0i}(\bm\gamma)] \log\log n})
\\\newtag
&\le
\lambda_d n^{\epsilon} + \mathcal{O}(\sqrt{n^{\epsilon}\log\log n}).
\end{align*}
  For $j=1,2$, denote
  \begin{align}\label{eq: def of Vji}
  V_{ji}(\bm p)&\equiv U_{1i}^j(\bm p)+(e^{{\bm\gamma}_0^\T \tilde{\bm x}_i}-1) U_{2i}^j(\bm p)\\
\label{eq: def of Wji}
  W_{ji}(\bm\gamma,\bm p)&\equiv U_{1i}^j(\bm p)+(e^{{\bm\gamma}^\T \tilde{\bm x}_i}-1) U_{2i}^j(\bm p)=V_{ji}(\bm p)+(e^{{\bm\gamma}^\T \tilde{\bm x}_i}-e^{{\bm\gamma}_0^\T \tilde{\bm x}_i}) U_{2i}^j(\bm p).
  \end{align}
Integration by parts implies
  \begin{align*}
    \E[V_{1i}(\bm p)]&=\E[U_{1i}(\bm p)+(e^{{\bm\gamma}_0^\T \tilde{\bm x}_i}-1) U_{2i}(\bm p)]\\&= \int_{\cal{X}}e^{{\bm\gamma}_0^\T\tilde{\bm x}}\int_0^\infty\left\{\frac{f_m(y|\bm x_0;\bm p)}{f(y|\bm x_0)}-1+(e^{{\bm\gamma}_0^\T\tilde{\bm x}}-1)\Biggr[
\frac{S_m(y|\bm x_0;\bm p)}{S(y|\bm x_0)}-1\Biggr]\right\} \\
&\quad\times
S(y|\bm x_0)^{e^{{\bm\gamma}_0^\T\tilde{\bm x}}-1} f(y|\bm x_0)dy dH(\bm x)  \\
&= \int_{\cal{X}}e^{{\bm\gamma}_0^\T\tilde{\bm x}}\int_0^\infty\Big\{ [f_m(y|\bm x_0;\bm p)- f(y|\bm x_0)]S(y|\bm x_0)^{e^{{\bm\gamma}_0^\T\tilde{\bm x}}-1}dy\\
&\quad -
  [S_m(y|\bm x_0;\bm p)- S(y|\bm x_0)]dS(y|\bm x_0)^{e^{{\bm\gamma}_0^\T\tilde{\bm x}}-1} \Big\}
 dH(\bm x) \\
&= - \int_{\cal{X}}e^{{\bm\gamma}_0^\T\tilde{\bm x}}  \Big[
  \{S_m(y|\bm x_0;\bm p)- S(y|\bm x_0)\}S(y|\bm x_0)^{e^{{\bm\gamma}_0^\T\tilde{\bm x}}-1}  \Big]_0^\infty
 dH(\bm x)\\
  &=0.
  \end{align*}
We also have
  \begin{align*}\label{eq: 2E(U1U2|x)}
2\E [U_{1i}(\bm p)U_{2i}(\bm p)|\bm x]
\newtag
&=-\E[ (e^{{\bm\gamma}_0^\T \tilde{\bm x}_i}-2) U_{2i}^2(\bm p)|\bm x_i].
  \end{align*}
  Therefore
 by (\ref{eq: 2E(U1U2|x)}) we have
  \begin{align*}
2\E [(e^{{\bm\gamma}_0^\T \tilde{\bm x}_i}-1) U_{1i}(\bm p)U_{2i}(\bm p)]
&=-\E[(e^{{\bm\gamma}_0^\T \tilde{\bm x}_i}-1) (e^{{\bm\gamma}_0^\T \tilde{\bm x}_i}-2) U_{2i}^2(\bm p)]
  \end{align*}
  and
  \begin{align*}
 \sigma^2[V_{1i}(\bm p)]&= \E[V^2_{1i}(\bm p)] =  \E\{[U_{1i}(\bm p)+(e^{{\bm\gamma}_0^\T \tilde{\bm x}_i}-1) U_{2i}(\bm p)]^2\}\\&=
\E [U_{1i}^2(\bm p)+(e^{{\bm\gamma}_0^\T \tilde{\bm x}_i}-1)^2 U_{2i}^2(\bm p)+2(e^{{\bm\gamma}_0^\T \tilde{\bm x}_i}-1) U_{1i}(\bm p)U_{2i}(\bm p)] \\
&=\E [U_{1i}^2(\bm p)+(e^{{\bm\gamma}_0^\T \tilde{\bm x}_i}-1) U_{2i}^2(\bm p)].
  \end{align*}
Thus
  \begin{align}\label{eq: E(V1^2)=Var(V1)=E(V2)}
\sigma^2[V_{1i}(\bm p)]
&={\E[V^2_{1i}(\bm p)] =\E[V_{2i}(\bm p)]=\chi_0^2(\bm p;\bm x_0)+\E [(e^{{\bm\gamma}_0^\T \tilde{\bm x}_i}-1) U_{2i}^2(\bm p)].}
  \end{align}
If $T$ is independent of covariate $\bm X$ then $\bm\gamma_0=\bm 0$ and
$\E[V^2_{1i}(\bm p)]= \chi_0^2(\bm p;\bm x_0).$
If $\bm\gamma_0\ne\bm 0$ we have $\bm\gamma\ne\bm 0$ for large $n$ and
\begin{equation}\label{eq: E[W2]=D0^2+D01^2}
\E[W_{2i}(\bm\gamma,\bm p)]=\chi_0^2(\bm p;\bm x_0)+D_{01}^2(\bm\gamma,\bm p;\bm x_0).
\end{equation}
Since $\bm\gamma^\T \bm x_0=\min\{\bm\gamma^\T\bm x: \bm x\in\mathcal{X}\}$, for any $\delta_0>0$ such that $\delta_0 e^{\delta_0}<1$, we have
\begin{align*}
D_{01}^2(\bm\gamma,\bm p;\bm x_0)&= \int_{\cal{X}} \int_0^\infty (e^{{\bm\gamma}^\T \tilde{\bm x}}-1)\Big[ \frac{S_m(y|\bm x_0;\bm p)}{S(y|\bm x_0)}-1\Big]^2  f(y|\bm x)dydH(\bm x)\\
&= \int_{{\bm\gamma}^\T \tilde{\bm x}\le \delta_0} \int_0^\infty (e^{{\bm\gamma}^\T \tilde{\bm x}}-1)\Big[ \frac{S_m(y|\bm x_0;\bm p)}{S(y|\bm x_0)}-1\Big]^2  f(y|\bm x)dydH(\bm x)\\
&~~~+\int_{{\bm\gamma}^\T \tilde{\bm x}> \delta_0} \int_0^\infty (e^{{\bm\gamma}^\T \tilde{\bm x}}-1)\Big[ \frac{S_m(y|\bm x_0;\bm p)}{S(y|\bm x_0)}-1\Big]^2  f(y|\bm x)dydH(\bm x)\\
&\ge \int_{{\bm\gamma}^\T \tilde{\bm x}\le \delta_0} \int_0^\infty (e^{{\bm\gamma}^\T \tilde{\bm x}}-1)\Big[ \frac{S_m(y|\bm x_0;\bm p)}{S(y|\bm x_0)}-1\Big]^2  f(y|\bm x)dydH(\bm x)\\
&~~~+\delta_0\int_{{\bm\gamma}^\T \tilde{\bm x}> \delta_0} \int_0^\infty \Big[ \frac{S_m(y|\bm x_0;\bm p)}{S(y|\bm x_0)}-1\Big]^2  f(y|\bm x)dydH(\bm x)\\
&= \int_{{\bm\gamma}^\T \tilde{\bm x}\le \delta_0} \int_0^\infty (e^{{\bm\gamma}^\T \tilde{\bm x}}-1-\delta_0)\Big[ \frac{S_m(y|\bm x_0;\bm p)}{S(y|\bm x_0)}-1\Big]^2  f(y|\bm x)dydH(\bm x)\\
&~~~+\delta_0\int_{\cal{X}} \int_0^\infty \Big[ \frac{S_m(y|\bm x_0;\bm p)}{S(y|\bm x_0)}-1\Big]^2  f(y|\bm x)dydH(\bm x).
\end{align*}
Hence we have
\begin{align*}
    D_{00}^2(\bm\gamma,\bm p; \bm x_0)
&\le \delta_0^{-1}D_{01}^2(\bm\gamma,\bm p; \bm x_0)- \int_{{\bm\gamma}^\T \tilde{\bm x}\le \delta_0} \int_0^\infty \\
&~~~\times\Big(\frac{e^{{\bm\gamma}^\T \tilde{\bm x}}-1}{\delta_0}-1\Big)\Big[ \frac{S_m(y|\bm x_0;\bm p)}{S(y|\bm x_0)}-1\Big]^2  f(y|\bm x)dydH(\bm x).
\end{align*}
Since for $x\ge 0$, $e^x-1\le xe^x$, we have, for ${\bm\gamma}^\T \tilde{\bm x}\le \delta_0$,
${\delta_0}^{-1}({e^{{\bm\gamma}^\T \tilde{\bm x}}-1})-1\le e^{{\bm\gamma}^\T \tilde{\bm x}}-1
\le \delta_0 e^{\delta_0} $. We have
\begin{align*}
    D_{00}^2(\bm\gamma,\bm p; \bm x_0)&\le \delta_0^{-1}D_{01}^2(\bm\gamma,\bm p; \bm x_0)\\
&~~~+\delta_0 e^{\delta_0}\int_{{\bm\gamma}^\T \tilde{\bm x}\le \delta_0} \int_0^\infty  \Big[ \frac{S_m(y|\bm x_0;\bm p)}{S(y|\bm x_0)}-1\Big]^2  f(y|\bm x)dydH(\bm x)\\
&\le \delta_0^{-1}D_{01}^2(\bm\gamma,\bm p; \bm x_0)+\delta_0 e^{\delta_0}D_{00}^2(\bm\gamma,\bm p; \bm x_0).
\end{align*}
Choosing $\delta_0$ to maximize $\delta_0(1-\delta_0e^{\delta_0})$,   we have
\begin{equation}\label{ineq: estimate of D00}
    D_{00}^2(\bm p; \bm x_0)\le \frac{\delta_0^{-1}}{1-\delta_0 e^{\delta_0}}D_{01}^2(\bm\gamma,\bm p; \bm x_0)< 5.59 D_{01}^2(\bm\gamma,\bm p; \bm x_0)
\end{equation}
and
\begin{align*}\label{ineq: estimate of D01^2(gamma0)}
    D_{01}^2(\bm\gamma_0,\bm p;\bm x_0)&=\E [|e^{{\bm\gamma}_0^\T \tilde{\bm x}_i}-1| U_{2i}^2(\bm p)]\\\newtag
&\le D_{01}^2(\bm\gamma,\bm p; \bm x_0)+ \E [|e^{{\bm\gamma}^\T \tilde{\bm x}_i}-e^{{\bm\gamma}_0^\T \tilde{\bm x}_i}| U_{2i}^2(\bm p)].
\end{align*}
By (\ref{eq: def of Wji})
\begin{align*}
  \sigma^2[W_{1i}(\bm\gamma,\bm p)] &=  \E[W_{2i}(\bm\gamma,\bm p)]+\{\E[W_{1i}(\bm\gamma,\bm p)]\}^2\\
 &~~~+ \E[(e^{{\bm\gamma}^\T \tilde{\bm x}_i}-1)(e^{{\bm\gamma}^\T \tilde{\bm x}_i}-e^{{\bm\gamma}_0^\T \tilde{\bm x}_i}) U_{2i}^2(\bm p) ].
\end{align*}
If $\E[W_{2i}(\bm\gamma,\bm p)]\le  n^{-1+\epsilon'}$, for any $\epsilon'\in(\epsilon,1/2)$, then by (\ref{eq: E(V1^2)=Var(V1)=E(V2)}), (\ref{eq: E[W2]=D0^2+D01^2}), (\ref{ineq: estimate of D01^2(gamma0)}) we have,  for all $\bm\gamma\in \mathbb{B}_{d}(n^{-1+\epsilon})$,
\begin{align*}\label{eq: esimate of E(W1i)}
\E[W_{1i}(\bm\gamma,\bm p)]&=\E[(e^{{\bm\gamma}^\T \tilde{\bm x}_i}-e^{{\bm\gamma}_0^\T \tilde{\bm x}_i})] U_{2i}(\bm p)\\\newtag
&=\mathcal{O}(n^{-1+(\epsilon+\epsilon')/2}) =o(n^{-1+\epsilon'}),\\
\label{eq: estimate of var(W1i)}
  \sigma^2[W_{1i}(\bm\gamma,\bm p)] &=  \E[W_{2i}(\bm\gamma,\bm p)]+\mathcal{O}(n^{-2+\epsilon+\epsilon'})+\mathcal{O}(n^{-3/2+\epsilon'+\epsilon/2})\\\newtag
&=\E[W_{2i}(\bm\gamma,\bm p)]+o(n^{-3/2+3\epsilon'/2}).
\end{align*}
For any $\epsilon'\in(\epsilon,1/2)$, if
\begin{equation}\label{eq: estimate of E(W2i)}
\E[W_{2i}(\bm\gamma,\bm p)]= \E[U^2_{1i}(\bm p)+(e^{{\bm\gamma}^\T \tilde{\bm x}_i}-1) U^2_{2i}(\bm p)]
= n^{-1+\epsilon'}
\end{equation}
then we have, by (\ref{eq: esimate of E(W1i)}),  (\ref{eq: estimate of var(W1i)}), and the LIL,
\begin{align*}
    \tilde{\mathcal{R}}_{01}({\bm\gamma},\bm p)&= -\sum_{i=1}^n  W_{1i}(\bm\gamma,\bm p) \\ &= -n\E[W_{1i}(\bm\gamma,\bm p)]+\mathcal{O}(\sqrt{n\sigma^2[W_{1i}(\bm\gamma,\bm p)]\log\log n})\\
&=o(n^{\epsilon'}),
\end{align*}
and, by  Kolmogorov's SLLN,
\begin{align*}
\tilde{\mathcal{R}}_{02}({\bm\gamma},\bm p)&=\frac{1}{2}\sum_{i=1}^n  W_{2i}(\bm\gamma,\bm p)
\\ &=\frac{n}{2} \E[W_{2i}(\bm\gamma,\bm p)] + o\{n\E[W_{2i}(\bm\gamma,\bm p)]\}.
\end{align*}
Thus, by (\ref{eq: log likelihood ratio}), there is an $\eta>0$ so that
 ${\mathcal{R}}({\bm\gamma},\bm p)=\sum_{j=0}^2\tilde{\mathcal{R}}_{0j}({\bm\gamma},\bm p) \ge \eta n^{\epsilon'} .$
While at $\bm p=\bm p_0$, $m\ge C_0n^{1/\rho}$,
 ${\mathcal{R}}({\bm\gamma},\bm p_0)=\mathcal{O}(n^{\epsilon})=o(n^{\epsilon'}).$
By (\ref{ineq: estimate of D01^2(gamma0)}), the minimizer $\tilde{\bm p}$ of ${\mathcal{R}}({\bm\gamma},\bm p)$ for the fixed $\bm\gamma$ satisfies
$D_0^2(\tilde{\bm p}; \bm x_0)\le \E[W_{2i}(\bm\gamma,\bm p)]+\E [|e^{{\bm\gamma}^\T \tilde{\bm x}_i}-e^{{\bm\gamma}_0^\T \tilde{\bm x}_i}| U_{2i}^2(\bm p)]
\le C' n^{-1+\epsilon'}$ for some constant $C'$ and $\tilde{\bm p}\in A_m(\epsilon_n)$.

Similarly, for any $\bm p$ that satisfies $D_0^2({\bm p}; \bm x_0)
\le C n^{-1+\epsilon}$, we can prove that the maximizer $\tilde{\bm\gamma}$ of
$\ell(\bm\gamma, \bm p)$  for the fixed $\bm p$ satisfies $\Vert\tilde{\bm\gamma}-\bm \gamma_0\Vert^2\le C' n^{-1+\epsilon'}$, for all $\epsilon'\in(\epsilon, 1/2)$, almost surely.
 The proof under condition \ref{C0} is complete.

\paragraph{Proof of Lemma \ref{lem: rate of convergence of gamma-tilde and fm(y|x0,p-tilde)} under condition \ref{C1}:}

Case I: current status data, all $\delta_i=1$. Let $G_1(\cdot|\bm x)$ be the conditional distribution of the censoring variable given $\bm X=\bm x$. We have
  \begin{align*}
    \E[U_{3i}(\bm\gamma,\bm p)]&= \E\Big\{ \Big[\frac{S_m(0|\bm x_0;\bm p)^{e^{{\bm\gamma}^\T \tilde{\bm X}}}-S_m(Y|\bm x_0;\bm p)^{e^{{\bm\gamma}^\T \tilde{\bm X}}}}{1-S(Y|\bm X)}  I(0 \le  T\le Y|\bm X) \\
&~~~+  \frac{S_m(Y|\bm x_0;\bm p)^{e^{{\bm\gamma}^\T \tilde{\bm X}}}-S_m(\infty|\bm x_0;\bm p)^{e^{{\bm\gamma}^\T \tilde{\bm X}}}}{S(Y|\bm X)} I(Y<T<\infty|\bm X)\Big]-1 \Big\}\\
&=  \int_{\cal{X}} \int_0^\infty
0 dG_1(y|\bm x)  dH(\bm x) =0,
  \end{align*}
  \begin{align*}\label{ineq: for var(U3i)=A_1^2(gamma,p)}
\E[U^2_{3i}(\bm\gamma,\bm p)]&=
 \int_{\cal{X}} \int_0^\infty  \frac{[S_m(y|\bm x;\bm\gamma,\bm p)-S(y|\bm x)]^2}{S(y|\bm x)[1-S(y|\bm x)]} dG_1(y |\bm x)  dH(\bm x)\\\newtag
&= \int_{\cal{X}}\int_0^\infty  \Big[\frac{S_m(y|\bm x_0; \bm p)^{e^{\bm\gamma^\T\tilde{\bm x}}}}{S(y|\bm x_0)^{e^{\bm\gamma_0^\T\tilde{\bm x}}}}-1\Big]^2O(y|\bm x) dG_1(y |\bm x)  dH(\bm x).
  \end{align*}
The LIL and the Kolmogorov's SLLN for $U_{3i}$'s implies, for all $\bm p\in \mathcal{A}(\epsilon_n)$,
  \begin{align*}\label{ineq: for var(U3i)=A_1^2(gamma,p)}
{\mathcal{R}}(\bm\gamma,\bm p)&=\tilde{\mathcal{R}}_{11}(\bm\gamma,\bm p)+\tilde{\mathcal{R}}_{12}(\bm\gamma,\bm p)\\
&=\mathcal{O}[\sigma(U_{3i})\sqrt{n\log\log n}]+n\sigma^2(U_{3i})+o[n\sigma^2(U_{3i})],~~~a.s..
  \end{align*}
By Taylor expansion,  with $u=e^{\bm\gamma^\T\tilde{\bm x}}$, $a=e^{\bm\gamma_0^\T\tilde{\bm x}}$,
$v=S_m(y|\bm x_0; \bm p)$, $b=S(y|\bm x_0)$,
\begin{equation}\label{eq: Taylor expansion of v^u/b^a}
\frac{v^u}{b^{a}}-1= (u-a) \log b+a \Big(\frac{v}{b}-1\Big)+R_2(\bm\gamma,\bm p),
\end{equation}
where
\begin{align*}
R_2(\bm\gamma,\bm p)
&=\frac{\bar b^{\bar a}}{2b^a}
\Big\{\Big[(\log \bar b)(u-a)+{\bar a} \frac{v-b}{\bar b}\Big]^2
+\Big[2(u-a)
- {\bar a} \frac{v-b}{\bar b}\Big]\frac{v-b}{\bar b}
\Big\},
\end{align*}
for some $(\bar a,\bar b)$ on the line segment joining $(u,v)$ and $(a,b)$, i.e.,
$$\bar a
=(1-\theta)e^{\bm\gamma_0^\T\tilde{\bm x}}+\theta e^{\bm\gamma^\T\tilde{\bm x}},\quad
\bar b=(1-\theta)S(y|\bm x_0)+\theta S_m(y|\bm x_0; \bm p),\quad 0\le\theta\le 1.$$
For all $\bm p\in A_m(\epsilon_n)$, $|v-b|/{b}\le \epsilon_n$,
\begin{align*}\label{eq: estimate of bb^ba/b^a}
\frac{\bar b^{\bar a}}{b^a}&=\frac{[b+\theta(v-b)]^{a+\theta(u-a)}}{b^a}
=\Big(1+\theta\frac{v-b}{b}\Big)^{a+\theta(u-a)} b^{\theta(u-a)}\\\newtag
&\le (1+\theta\epsilon_n)^{a+\theta(u-a)}b^{\theta(u-a)}\le C(1-|u-a|\log b ).
\end{align*}
For $k=1,2$,
\begin{equation}\label{eq: estimate of bb^k/b^k, k=1,2}
\frac{b^k}{\bar b^{k}}=\frac{b^k}{[b+\theta(v-b)]^{k}}
=\Big(1+\theta\frac{v-b}{b}\Big)^{-k} \le (1-\theta\epsilon_n)^{-k} \le C'.
\end{equation}
Since $\log(1+z)=\sum_{k=1}^\infty (-1)^{k+1} \frac{z^{k}}{k}$, $|z|<1$, we have, for all $\bm p\in A_m(\epsilon_n)$,
\begin{equation}\label{eq: expansion of log (b+eps)}
\log\bar b=\log[b+\theta(v-b)]=\log b
+\mathcal{O} (|{v-b}|/{b})=\log b+\mathcal{O} (\epsilon_n).
\end{equation}
For all positive integer $k$ we have 
\begin{equation}\label{ineq: |z(log z)^k|<= k^k e^(-k)}
|z(\log z)^k|\le k^k e^{-k},\quad z\in[0,1].
\end{equation}
For any $\bm\gamma\in \mathbb{B}_{d}(n^{-1+\epsilon})$, $\epsilon\in(0,1/2)$ and $\bm x_0$ such that $\bm\gamma^\T\bm x_0=\max_{\bm x\in\mathcal{X}}\bm\gamma^\T\bm x$.
If, for $\epsilon'\in (\epsilon,1/2)$,
$$D_1^2(\bm p;\bm x_0)=\int_{\cal{X}} \int_0^\infty  \Big[\frac{S_m(y|\bm x_0; \bm p)}{S(y|\bm x_0)}-1\Big]^2O(y|\bm x)  dG_1(y |\bm x)  dH(\bm x)=n^{-1+\epsilon'}$$
then it follows from (\ref{eq: Taylor expansion of v^u/b^a}--\ref{ineq: |z(log z)^k|<= k^k e^(-k)}),   the triangular inequality, and inequality $|u(\log u)^k|\le k^k e^{-k}$, $u\in[0,1]$, for positive integer $k$, that, for all $\bm p\in A_m(\epsilon_n)$,
\begin{align*}\label{ineq: sigma^2(U3) estimated by Mikovski ineq}
\sigma^2(U_{3i})&\ge \left|D_1(\bm p;\bm x_0) -\Big\{\int_{\cal{X}}\int_0^\infty  \Big[e^{\bm\gamma_0^\T\tilde{\bm x}} (e^{\bm\gamma^\T\tilde{\bm x}-\bm\gamma_0^\T\tilde{\bm x}}-1) \log S(y|\bm x_0) \Big]^2  \right.\\\newtag
&~~~\left. O(y|\bm x)  dG_1(y |\bm x)  dH(\bm x)\Big\}^{1/2}\right|^2 +o(n^{-1+\epsilon'}),~~~a.s..
\end{align*}
 By  (\ref{ineq: |z(log z)^k|<= k^k e^(-k)}), 
$\sigma^2(U_{3i})\ge |n^{-(1-\epsilon')/2}-2e^{-1}\E^{1/2}[O(Y|\bm X)]\mathcal{O}(n^{-(1-\epsilon)/2})|^2+o(n^{-1+\epsilon'}).$
Thus, there is an $\eta_0>0$, so that, for all $\bm p$ that satisfy $D_1^2(\bm p;\bm x_0) =n^{-1+\epsilon'}$, we have
$\mathcal{R}(\bm \gamma,\bm p)\ge \eta_0  n^{\epsilon'}$, a.s..
At $\bm p=\bm p_0$, with $m\ge C_0n^{1/\rho}$,
 $\mathcal{R}(\bm \gamma,\bm p_0)=\mathcal{O}(n^{\epsilon})$, a.s..
Therefore $\mathcal{R}(\bm \gamma,\bm p)$ is minimized by $\tilde{\bm p}=\tilde{\bm p}(\bm\gamma)$ such that
\begin{equation}\label{eq: D1^2 <log n/n}
D_1^2(\tilde{\bm p};\bm x_0)=\int_{\cal{X}}\int_0^\infty  \Big[\frac{S_m(y|\bm x_0; \tilde{\bm p})}{S(y|\bm x_0)}-1\Big]^2O(y|\bm x)  dG_1(y |\bm x)  dH(\bm x)<n^{-1+\epsilon'}.
\end{equation}
Similarly, by (\ref{ineq: sigma^2(U3) estimated by Mikovski ineq}), if $D_1^2(\tilde{\bm p};\bm x_0)<n^{-1+\epsilon}$ for an $\bm x_0\in\mathcal{X}$, then the minimizer $\tilde{\bm\gamma}=\tilde{\bm\gamma}(\bm p)$ of $\mathcal{R}(\bm \gamma,\bm p)$ satisfies
$\tilde{\bm\gamma}\in\mathcal{B}_{d}(n^{-1+\epsilon'})$ for all $\epsilon'\in (\epsilon,1/2)$.

\paragraph{Proof of Lemma \ref{lem: rate of convergence of gamma-tilde and fm(y|x0,p-tilde)} under condition \ref{C2}:}
For Case II 
interval censored data $\delta_i=1$, let $G_2(y_1,y_2|\bm x)$ be the conditional distribution of $(Y_1,Y_2)$ given $\bm X=\bm x$. We have
  \begin{align*}
    \E[U_{3i}(\bm\gamma,\bm p)]&= \E\Big\{ \Big[\frac{S_m(0|\bm x_0;\bm p)^{e^{{\bm\gamma}^\T \tilde{\bm X}}}-S_m(Y_1|\bm x_0;\bm p)^{e^{{\bm\gamma}^\T \tilde{\bm X}}}}{1-S(Y_1|\bm X)}  I(0 {\le} T\le Y_1|\bm X) \\
&~~~+  \frac{S_m(Y_1|\bm x_0;\bm p)^{e^{{\bm\gamma}^\T \tilde{\bm X}}}-S_m(Y_2|\bm x_0;\bm p)^{e^{{\bm\gamma}^\T \tilde{\bm X}}}}{S(Y_1|\bm X)-S(Y_2|\bm X)}  I(Y_1<T\le Y_2|\bm X) \\
&~~~+  \frac{S_m(Y_2|\bm x_0;\bm p)^{e^{{\bm\gamma}^\T \tilde{\bm X}}}-S_m(\infty|\bm x_0;\bm p)^{e^{{\bm\gamma}^\T \tilde{\bm X}}}}{S(Y_2|\bm X)} I(Y_2<T<\infty|\bm X)\Big]-1 \Big\}\\
&=  \int_{\cal{X}} \int_0^\infty\int_0^{y_2}
0 dG_2(y_1,y_2|\bm x)  dH(\bm x) =0.
  \end{align*}
Similarly
  \begin{align*}
   \E[U^2_{3i}(\bm\gamma,\bm p)]
&=\int_{\cal{X}}\mathop{\int\int}_{0<y_1<y_2<1}\sum_{i=1}^3  \Big[\frac{S_m(y_{i-1}|\bm x;\bm\gamma,\bm p)-S_m(y_{i}|\bm x;\bm\gamma,\bm p)}{S(y_{i-1}|\bm x)-S(y_i|\bm x)}-1\Big]^2\\
  &\quad\quad\quad\times[S(y_{i-1}|\bm x)-S(y_i|\bm x)] g(y_1,y_2)dy_1dy_2 dH(\bm x).
  \end{align*}
Simplifying notations  $\tilde S_i=S_m(Y_{i}|\bm X;\bm\gamma,\bm p)$, $S_i=S(Y_i|\bm X)$, and $\Lambda_i=\Lambda(Y_i|\bm X)$, $i=1,2,$
we have, clearly,
  \begin{align*}
   \E[U^2_{3i}(\bm\gamma,\bm p)]
 & \ge \E\Big\{\frac{(\tilde S_1-S_1)^2}{1-S_1}  + \frac{(\tilde S_2-S_2)^2}{S_2} \Big\}\\
&=\E\Big\{\Big(\frac{\tilde S_1}{S_1}-1\Big)^2\frac{S_1^2}{1-S_1}  + \Big(\frac{\tilde S_2}{S_2}-1\Big)^2 {S_2} \Big\}\\
&=\E\Big\{\Big[\frac{S_m(Y_{1}|\bm X;\bm\gamma,\bm p)}{S(Y_1|\bm X)}-1\Big]^2O(Y_1|\bm X)S(Y_1|\bm X)\Big\}\\
&~~~+\E\Big\{\Big[\frac{S_m(Y_{2}|\bm X;\bm\gamma,\bm p)}{S(Y_2|\bm X)}-1\Big]^2 S(Y_2|\bm X)\Big\}.
  \end{align*}
Thus the proof under condition \ref{C2} can be done by the argument similar to the proof under condition \ref{C1}.
The proof of Lemma \ref{lem: rate of convergence of gamma-tilde and fm(y|x0,p-tilde)} is complete.

Now we prove Theorem \ref{thm: rate of convergence of mable (gamma-hat, p-hat)}.
Let $\mathbb{B}_{d}(r)=\{\bm\gamma: \Vert \bm\gamma-\bm\gamma_0 \Vert\le r\}$, where  $\Vert\cdot\Vert$ denotes the
Euclidean norm in $R^d$.  For a decreasing positive sequence $\epsilon_n\searrow 0$ slowly as $n\to\infty$, e.g., $\epsilon_n=1/\log (n+2)$, let $A_m(\epsilon_n)$ be a subset of $\mathbb{S}_{{m^*}}$ so that, for all  $t\in [0,b]$, $|f_m(t|\bm x_0;\bm p)-f(t|\bm x_0)|/f(t|\bm x_0)\le\epsilon_n$. Clearly, for all $\bm p\in A_m(\epsilon_n)$, we have
 $|S_m(t|\bm x_0;\bm p)-S(t|\bm x_0)|/S(t|\bm x_0)\le\epsilon_n$.

If $\bm\gamma^{(0)}$ is chosen to be an efficient and asymptotically normal estimator of $\bm\gamma$ as in \cite{Cox1972} and \cite{Huang-and-Wellner-1997}, then, under the conditions of the theorem, for large $n$,  almost surely
$\Vert \bm\gamma^{(0)}-\bm\gamma_0\Vert^2<n^{-1+\epsilon}.$
Lemma \ref{lem: rate of convergence of gamma-tilde and fm(y|x0,p-tilde)} and the convergence of $(\bm\gamma^{(s)},\bm p^{(s)})$ imply that $\Vert\hat{\bm\gamma}-\bm\gamma_0\Vert\le n^{-1+\epsilon}$, $D^2_i(\hat{\bm p};\hat{\bm x}_0)\le n^{-1+\epsilon}$, and $\hat{\bm p}\in A_m(\epsilon_n)$. 
The proof is complete.
\subsection{Proof of Theorem \ref{thm: Asym normality of gamma-tilde}.}

\paragraph{Uncensored Data:  all $\delta_i=0$}
Expansion of $Q(\tilde{\bm\gamma},  S_m)=\frac{\partial\ell_m(\tilde{\bm\gamma},  {\bm p}_0)}{\partial\bm\gamma}$ at $\bm\gamma_0$:
\begin{align*}
\bm 0&=n^{-1/2}Q(\tilde{\bm\gamma},  S_m)=\bm Z_n - \bm J_n \sqrt{n}(\tilde{\bm\gamma}-\bm\gamma_0)+n^{-1/2}R_n(\tilde{\bm \gamma}),
\end{align*}
where
\begin{align*}
\bm Z_n & =n^{-1/2}\sum_{i=1}^n [1+e^{\bm\gamma_0^\T \tilde{\bm x}_i} \log S_m(y_i|\bm x_0;{\bm p}_0)]{\tilde{\bm x}_i}\\
\bm J_n
&= -\frac{1}{n}\sum_{i=1}^n e^{\bm\gamma_0^\T \tilde{\bm x}_i}\log S_m(y_i|\bm x_0;{\bm p}_0){\tilde{\bm x}_i}{\tilde{\bm x}_i}^\T,
\\
   R_n(\tilde{\bm \gamma})&=\frac{1}{2}\sum_{i=1}^ne^{\bar{\bm\gamma}^\T \tilde{\bm x}_i} \log S_m(y_i|\bm x_0;{\bm p}_0)[(\tilde{\bm \gamma}-\bm\gamma_0)^\T{\tilde{\bm x}_i}]^2 {\tilde{\bm x}_i},
\end{align*}
and $\bar{\bm\gamma}={\bm\gamma}_0+\theta(\tilde{\bm \gamma}-{\bm\gamma}_0)$
 for some $\theta\in [0,1]$.
If $m=m_n$ satisfies $n^{1/2}m^{-\rho/2}=o(1)$ then
$$\bm J_n\to -\E[\log S(T|\bm X) \tilde{\bm X}\tilde{\bm X}^\T]=\E(\tilde{\bm X}\tilde{\bm X}^\T)=\mathcal{I}$$ and $\bm Z_n$ converges in distribution to normal with mean $\bm 0$ and variance
$\mathcal{I}$.
For any $\epsilon>0$ and large $n$, $R_n(\tilde{\bm \gamma})=\mathcal{O}(n^{\epsilon})$, a.s.. Thereofor
$\sqrt{n}(\tilde{\bm\gamma}-\bm\gamma_0)= \bm J_n^{-1}[\bm Z_n +\mathcal{O}(n^{-1/2+\epsilon})]$
 converges in distribution to normal with mean $\bm 0$ and variance
$\mathcal{I}^{-1}$.
\paragraph{Interval censored Data: all $\delta_i=1$}
Expansion of $Q(\tilde{\bm\gamma},  S_m)=\frac{\partial\ell_m(\tilde{\bm\gamma},  {\bm p}_0)}{\partial\bm\gamma}$ at $\bm\gamma_0$ gives
\begin{align*}
\bm 0&=n^{-1/2}Q(\tilde{\bm\gamma},  S_m)=\bm Z_n - \bm J_n \sqrt{n}(\tilde{\bm\gamma}-\bm\gamma_0)+n^{-1/2}R_n(\tilde{\bm \gamma}),
\end{align*}
where 
\begin{align*}
\bm Z_n & =n^{-1/2}\sum_{i=1}^n
 \frac {\dot S_m(y_{i1}|\bm x_i;\bm\gamma_0;\bm p_0)-\dot S_m(y_{i2}|\bm x_i;\bm\gamma_0;\bm p_0)}{S_m(y_{i1}|\bm x_i;\bm\gamma_0;\bm p_0)-S_m(y_{i2}|\bm x_i;\bm\gamma_0;\bm p_0)},\\
\bm J_n
&= -\frac{1}{n}\sum_{i=1}^n \Big\{
\frac {\ddot S_m(y_{i1}|\bm x_i;\bm\gamma_0;\bm p_0)-\ddot S_m(y_{i2}|\bm x_i;\bm\gamma_0;\bm p_0)}{S_m(y_{i1}|\bm x_i;\bm\gamma_0;\bm p_0)-S_m(y_{i2}|\bm x_i;\bm\gamma_0;\bm p_0)}
 \\
  &~~~~- \frac {[\dot S_m(y_{i1}|\bm x_i;\bm\gamma_0;\bm p_0)-\dot S_m(y_{i2}|\bm x_i;\bm\gamma_0;\bm p_0)]^{\otimes 2}}{[S_m(y_{i1}|\bm x;\bm\gamma_0;\bm p_0)-S_m(y_{i2}|\bm x_i;\bm\gamma_0;\bm p_0)]^2} \Big\},
\end{align*}
for any $\epsilon>0$ and large $n$, $R_n(\tilde{\bm \gamma})=\mathcal{O}(n^{\epsilon})$, a.s..
If $m=m_n$ satisfies $n^{1/2}m^{-\rho/2}=o(1)$ then, for current status data,
\begin{align*}
\bm J_n &\to -\E\Big\{\Big[-S(Y|\bm X)\Lambda^2(Y|\bm X)+S(Y|\bm X)\Lambda^2(Y|\bm X) \Big]\tilde{\bm X}\tilde{\bm X}^\T \Big\}\\
&~~~~+ \E\Big\{\Big[\frac{ S^2(Y|\bm X)\Lambda^2(Y|\bm X)}{1-S(Y|\bm X)}+\frac{S^2(Y|\bm X)\Lambda^2(Y|\bm X)}{S(Y|\bm X)} \Big]\tilde{\bm X}\tilde{\bm X}^\T \Big\}\\
&= \E\Big\{\Big[O(Y|\bm X)\Lambda^2(Y|\bm X) \Big]\tilde{\bm X}\tilde{\bm X}^\T \Big\}\equiv \mathcal{I}
\end{align*}
and for Case $k$ ($k\ge 2$) interval censored data,
\begin{align*}
\bm J_n &\to -\E\Big\{\Big[-S(Y_1|\bm X)\Lambda^2(Y_1|\bm X)+S(Y_1|\bm X)\Lambda^2(Y_1|\bm X)-S(Y_2|\bm X)\Lambda^2(Y_2|\bm X)\\
&~~~+S(Y_2|\bm X)\Lambda^2(Y_2|\bm X) \Big]\tilde{\bm X}\tilde{\bm X}^\T \Big\}\\
&+ \E\Big\{\Big[\frac{ S^2(Y_1|\bm X)\Lambda^2(Y_1|\bm X)}{1-S(Y_1|\bm X)}+\frac{[S(Y_1|\bm X)\Lambda(Y_1|\bm X)-S(Y_2|\bm X)\Lambda(Y_2|\bm X)]^2}{S(Y_1|\bm X)-S(Y_2|\bm X)}\\
 &~~~+\frac{S^2(Y_2|\bm X)\Lambda^2(Y_2|\bm X)}{S(Y_2|\bm X)} \Big]\tilde{\bm X}\tilde{\bm X}^\T \Big\}\\
&= \E\Big\{\Big[\frac{ (S_1\Lambda_1)^2(1-S_2)}{(1-S_1)(S_1-S_2)}-2\frac{ (S_1\Lambda_1)(S_2\Lambda_2)}{S_1-S_2}+\frac{ (S_2\Lambda_2)^2S_1}{S_2(S_1-S_2)} \Big]\tilde{\bm X}\tilde{\bm X}^\T \Big\}
\equiv \mathcal{I},
\end{align*}
where $\Lambda_i=\Lambda(Y_i|\bm X)$, $i=1,2$. It is clear
\begin{align*}
\mathcal{I} &\ge \E\Big\{\Big[\frac{ S^2(Y_1|\bm X)\Lambda^2(Y_1|\bm X)}{1-S(Y_1|\bm X)}+ \frac{S^2(Y_2|\bm X)\Lambda^2(Y_2|\bm X)}{S(Y_2|\bm X)} \Big]\tilde{\bm X}\tilde{\bm X}^\T \Big\}\\
&= \E\Big\{\Big[O(Y_1|\bm X)S(Y_1|\bm X)\Lambda^2(Y_1|\bm X) +  S(Y_2|\bm X)\Lambda^2(Y_2|\bm X) \Big]\tilde{\bm X}\tilde{\bm X}^\T \Big\}.
\end{align*}

In both cases, $\bm Z_n$ converges in distribution to normal with mean $\bm 0$ and variance
$\mathcal{I}$.
For any $\epsilon>0$ and large $n$, $R_n(\tilde{\bm \gamma})=\mathcal{O}(n^{\epsilon})$, a.s.. Hence
$\sqrt{n}(\tilde{\bm\gamma}-\bm\gamma_0)= \bm J_n^{-1}[\bm Z_n +\mathcal{O}(n^{-1/2+\epsilon})]$
 converges in distribution to normal with mean $\bm 0$ and variance
$\mathcal{I}^{-1}$.
\bibliographystyle{apalike}
\def\cprime{$'$} \def\cprime{$'$}
  \def\polhk#1{\setbox0=\hbox{#1}{\ooalign{\hidewidth
  \lower1.5ex\hbox{`}\hidewidth\crcr\unhbox0}}} \def\cprime{$'$}

 \end{document}